\documentclass[prc,aps,superscriptaddress,showpacs,twocolumn]{revtex4-2}
\usepackage{hyperref}
\usepackage{graphicx}
\usepackage{amsfonts}
\usepackage{amssymb,amsbsy}
\usepackage{dcolumn}
\usepackage{ulem}
\usepackage{bm}
\usepackage{xcolor}
\usepackage[utf8]{inputenc}
\usepackage{subeqnarray}
\usepackage{bm} 
\newcommand{\bwt}{\begin{widetext}}
\newcommand{\ewt}{\end{widetext}}
\newcommand{\beq}{\begin{equation}}
\newcommand{\eeq}{\end{equation}}
\newcommand{\bea}{\begin{eqnarray}}
\newcommand{\eea}{\end{eqnarray}}

\title{Collective modes and magnetic field}
\date{\today}

\begin{document}
\title{Longitudinal collective modes in relativistic asymmetric magnetized nuclear matter within the covariant Vlasov approach}
\author{Aziz Rabhi} 
\email{rabhi@uc.pt}
\affiliation{IPEST La Marsa, Carthage University, Tunisia.}
\affiliation{CFisUC, Department of Physics, University of Coimbra, 3004-516 Coimbra, Portugal.}
\author{Olfa Boukari} 
\email{olfa.boukari@isepbg.ucar.tn}
\affiliation{ISEP-BG La Soukra, Carthage University, Tunisia.}
\author{Sidney S. Avancini} 
\email{avancini@ufsc.br}
\affiliation{Departamento de F\'{\i}sica, Universidade Federal de Santa Catarina, Florian\'opolis, SC, CP. 476, CEP 88.040-900, Brazil.} 
\author{Constan\c ca Provid\^encia}
\email{cp@uc.pt}
\affiliation{CFisUC, Department of Physics, University of Coimbra, 3004-516 Coimbra, Portugal.}

\begin{abstract}
The neutron-proton-electron (npe) matter under strong magnetic field is studied in the context of the covariant Vlasov approach. A covariant relativistic approach based on the Vlasov equation is applied to the study of infinite asymmetric magnetized nuclear matter. We use several relativistic mean-field nuclear models with non-linear terms. The dispersion relations for the longitudinal modes are obtained, and the isovector and isoscalar collective modes are determined in a wide range of densities as a function of the isospin asymmetry, momentum transfer, and magnetic field. A strong magnetic field gives rise to the appearance of low-lying isovector modes that propagate in nuclear matter, not present in non-magnetized matter. Neutron-like modes are essentially not affected by the presence of a strong magnetic field. In the presence of a strong magnetic field, Landau quantization modifies the proton-like collective modes, leading to the emergence of new branches associated with distinct Landau levels. These new modes can propagate even at high densities and exhibit isoscalar or isovector character.
\end{abstract}
\maketitle

\section{Introduction}

Understanding the properties of isospin-rich nuclear matter is essential for modeling a wide range of astrophysical environments, including the interiors of compact stars, supernova cores, and neutron stars. This endeavor requires a multidisciplinary theoretical framework that brings together astrophysics, nuclear many-body theory, and thermodynamics. In particular, dense nuclear matter subjected to extreme conditions, such as high baryon density, large neutron-proton asymmetry, and intense magnetic fields, exhibits complex dynamical and structural behavior that can significantly impact macroscopic observables. Recent advances in both observations and theoretical modeling have underscored the need for detailed studies of how external fields and microphysical interactions shape the response and stability of matter in these regimes.

The study of nuclear matter under the influence of strong magnetic fields is of increasing importance, particularly in the context of compact astrophysical objects such as neutron stars and, more specifically, magnetars. These highly magnetized stars are believed to host magnetic fields ranging from $10^{14}$ to $10^{15}$G on the surface~\cite{Duncan1992, Duncan1995}, and potentially exceeding $10^{18}$G in their interiors. Such extreme fields are expected to significantly modify the microscopic properties and macroscopic behavior of dense matter, affecting not only the equation of state (EOS), but also the transport properties and dynamical response of the system. Understanding these effects is essential for modeling neutron star oscillations, cooling, magnetic field decay~\cite{Pons2013}, and other phenomena relevant to astrophysical observations, including gravitational waves and X-ray emissions.

One of the key tools for probing the dynamical behavior of nuclear matter is the study of collective modes. These modes, which arise from coherent oscillations of the system’s constituents, provide insight into the stability, compressibility, and structure of the interaction of dense matter. In particular, the behavior of isoscalar and isovector collective modes can reveal the nature of nuclear interactions and their sensitivity to various external conditions, such as temperature, asymmetry, and magnetic fields. We consider that isovector modes correspond to oscillations in which protons and neutrons move out of phase, whereas isoscalar modes involve in-phase motion of the two nucleon species. In the presence of a magnetic field, the charged components of nuclear matter - most notably the protons - experience Landau quantization, altering their density of states and modifying their response to perturbations. This, in turn, can affect the collective oscillations of both charged and neutral particles through the self-consistent mean fields.

To account for dynamical effects, we employ the covariant Vlasov approach, which is the semi-classical limit of the relativistic transport equation. This method allows for a fully covariant and self-consistent treatment of the nuclear response, incorporating the meson fields and electromagnetic interactions in a unified manner. We extend previous studies employing the covariant Vlasov approach~\cite{avancini2018, rabhi2025}, focusing on the investigation of collective modes in relativistic asymmetric magnetized nuclear matter. Our study follows earlier work in which the stability of npe matter under strong magnetic fields was examined within the same framework~\cite{avancini2018}, or collective modes in cold asymmetric nuclear matter without magnetic fields were discussed~\cite{rabhi2025}. These efforts are pursued by incorporating magnetic field effects, to study the collective modes in magnetized nuclear matter within the covariant Vlasov approach. We shall restrict ourselves to longitudinal modes and investigate only the nuclear collective modes. Our analysis is performed within the framework of relativistic mean-field (RMF) theory, which has proven successful in describing the bulk properties of nuclear matter and finite nuclei.

We focus on cold nuclear matter composed of neutrons and protons (np matter), considering various proton fractions relevant for both symmetric and asymmetric systems. The ground state of the system in the presence of the magnetic field is constructed using the RMF equations of motion, accounting for the Landau-level structure of protons while neglecting the anomalous magnetic moment (AMM) of nucleons in this initial analysis. Small-amplitude oscillations around this magnetized background are then studied by linearizing the Vlasov equation. The resulting dispersion relations describe the propagation of longitudinal collective modes with a fixed wave vector q parallel to the magnetic field.

Special attention is given to the behavior of isoscalar-like and isovector-like modes. We analyze how the magnetic field modifies the speed of sound in these modes and how the coupling between neutrons and protons through the meson fields leads to indirect magnetic effects on the neutron sector, despite the absence of a direct coupling to the magnetic field. In particular, we show that the proton response is strongly altered by Landau quantization, which feeds back into the neutron dynamics via the isovector meson field. In contrast, the isoscalar-like neutron modes are only weakly affected, due to the nearly symmetric nature of the proton response in these in-phase oscillations.

This study provides a detailed characterization of the magnetic field-induced modifications of the collective excitations in nuclear matter, highlighting both the direct and indirect effects of the magnetic field on the system. The formalism developed here can be extended to include finite-temperature, beta-equilibrated matter, anomalous magnetic moments, and the presence of leptonic or hyperonic components. Extensions of the present formalism to transverse modes, are also possible and will be considered in future studies. Such extensions are essential for a comprehensive understanding of matter under the extreme conditions encountered in neutron stars and other compact objects.

In Sec. II we show a brief review of the Covariant Vlasov equation approach for nuclear neutral matter, including electrons and the electromagnetic field, and in the presence of an external magnetic field. The numerical results are presented and discussed in Sec. III and the conclusions are drawn in Sec. IV.
%
%
%
%
%
%
%
%
%
\section{Formalism}
The present work is formulated within a covariant density functional theory, which provides a relativistic description of nuclear systems based on effective Lagrangians. In this approach, the interaction between nucleons is mediated by the exchange of mesons and the photon field. Specifically, we consider the scalar-isoscalar $\sigma$ meson ($\phi$), the vector-isoscalar $\omega$ meson ($V^\mu$), the vector-isovector $\rho$ meson ($\vec{b}^\mu$), and the electromagnetic field ($A^\mu$). These fields are characterized by their respective masses $m_s$, $m_v$, and $m_\rho$, and their couplings to the nucleons are denoted by $g_s$, $g_v$, $g_\rho$, and $e$, respectively.

The Lagrangian density of the system, written in natural units ($\hbar = c = 1$), is given by
\begin{equation}
{\cal L}=\sum_{j=p,n,e} {\cal L}_j + \cal L_\sigma + {\cal L}_\omega +
{\cal L}_\rho + {\cal L}_{\omega \rho } + {\cal L}_{A} \ ,
\label{lag}
\end{equation}
with
\begin{equation}
{\cal L}_j=\bar \psi^{(j)}\left[\gamma^\mu i D_\mu^{(j)}-M^*_j \right]\psi^{(j)} ,\nonumber
\end{equation}
where the covariant derivative is defined as $ iD^{(j)}_{\mu}~= ~ i \partial_{\mu} - 
{\cal V}^{(j)}_{\mu} $, the spinor $\psi$ describes a single nucleon, a proton or a neutron and the index $(j)=(n,p,e)$ stands for the neutron, proton, and electron,
\begin{equation}
 {\cal V}^{(j)}_{\mu} = 
 \left\{
       \begin{array}{l}
         g_v V_\mu  + \frac{g_\rho}{2}\, \vec{b}_\mu+ e\, A_\mu \ ~ ,~j=p \\  
         g_v {V}_\mu -\frac{g_\rho}{2}\, \vec{b}_\mu \ ~ ,~j=n \\
         - e\, A_\mu \ ~ ,~ j=e
     \end{array}    \right. \ ,  
\end{equation}
where $M^*_p=M^*_n=M^*=M-g_s\phi$, $M^*_e=m_e$, $e=\sqrt{4\pi/137}$ is the electromagnetic coupling constant.
In the present study, we consider the mean-field approximation. We employ the RMF models NL3~\cite{nl3}, NL3~$\omega\rho$~\cite{Pais16,Horowitz01}, and FSU~\cite{fsu} with the constant couplings listed in Table~\ref{table1}. The  properties of these models are summarized in Table~\ref{tab:nuclear}. 
This set of models is chosen because they span a quite large range of nuclear matter parameters, in particular incompressibility, symmetry energy, and its slope and effective mass, all at saturation.

\begin{table*}
\caption{Parameter sets of the RMF models NL3, NL3 $\omega\rho$, and FSU used in this study.}
\label{table1}
\begin{ruledtabular}
\begin{tabular}{c c c c c c c c c c c cc}
&$\rho_{0}$  & $m_{s}$ &$m_{v}$&$m_{\rho}$& $g_s$  & $ g_v$ & $g_\rho$ & $\kappa$& $\lambda$ & $\xi$ & $\Lambda_{v}$ \\
&($fm^{-3}$) & MeV  & MeV & MeV &  &  & &  & & & \\
\hline
NL3             & 0.148  & 508.194 & 782.501 & 763.000 & 10.217 & 12.868 &  8.948 & 4.3840 & -173.31 & 0.0000 & 0.0000\\
NL3$\omega\rho$ & 0.148  & 508.194 & 782.501 & 763.000 & 10.217 & 12.868 & 11.277 & 4.3840 & -173.31 & 0.0000 & 0.0300 \\
FSU             & 0.1505 & 491.500 & 782.500 & 763.000 & 10.592 & 14.302 & 11.767 & 1.7976 &  299.13 & 0.0600 & 0.0300 \\
\end{tabular}
\end{ruledtabular}
\end{table*}

\begin{table}[h!]
\caption{Nuclear matter properties of the RMF models considered in this study \label{tab:nuclear}}
\begin{ruledtabular}
 \begin{tabular}{ccccccc}
  & $n_0$ & $M^*/M$ & $B$ & $K$ & $E_\mathit{sym}$   & $L$  \\
  & $ [\mathrm{fm}^{-3}]$ &  & [MeV]&[MeV]&[MeV]& [MeV]\\\hline 
NL3 &0.148& 0.596&-16.24&271&37.4&118 \\
NL3$\omega\rho$&0.148& 0.596 &-16.24&271&31.5&55\\
FSU & 0.148 & 0.611 &-16.30 & 230 & 32.59 & 60.5  \\
\end{tabular}
\end{ruledtabular}
\end{table}
The contributions of the meson and electromagnetic fields to the Lagrangian density, as introduced in Eq.~(\ref{lag}), are expressed as follows:
\begin{eqnarray}
\mathcal{L}_{{\sigma }} &=&\frac{1}{2}\left( \partial _{\mu }\phi \partial %
^{\mu }\phi -m_{s}^{2}\phi ^{2}-\frac{1}{3}\kappa \phi ^{3}-\frac{1}{12}%
\lambda \phi ^{4}\right) \ , \nonumber \\
\mathcal{L}_{{\omega }} &=&\frac{1}{2} \left(-\frac{1}{2} \Omega _{\mu \nu }
\Omega ^{\mu \nu }+ m_{v}^{2}V_{\mu }V^{\mu }
+\frac{1}{12}\xi g_{v}^{4}(V_{\mu}V^{\mu })^{2} 
\right) \ , \nonumber \\
\mathcal{L}_{{\rho }} &=&\frac{1}{2} \left(-\frac{1}{2}
{\vec{B}}_{\mu \nu }\cdot {\vec{B}}^{\mu
\nu }+ m_{\rho }^{2} \vec{b}_{\mu }\cdot \vec{b}^{\mu } \right)   \ , \nonumber \\
\mathcal{L}_{\omega \rho } &=& \Lambda_v g_v^2 g_\rho^2 V_{\mu }V^{\mu }
\vec{b}_{\nu }\cdot \vec{b}^{\nu }   \nonumber \\
\cal{L}_{A} &=&-\frac{1}{4} F_{\mu \nu }F ^{\mu \nu }~, \label{mesonlag}
\end{eqnarray}
where $\Omega _{\mu \nu }=\partial _{\mu }V_{\nu }-\partial _{\nu }V_{\mu }$, 
$\vec{B}_{\mu \nu }=\partial _{\mu }\vec{b}_{\nu }-\partial _{\nu }
\vec{b}
_{\mu }-\Gamma_{\rho }(\vec{b}_{\mu }\times \vec{b}_{\nu })$ and 
$F_{\mu \nu }=\partial _{\mu }A_{\nu }-\partial _{\nu }A_{\mu }$.
The parameters $\kappa$, $\lambda$, and $\xi$ are self-interacting couplings and the $\omega-\rho$ coupling $\Lambda_v$ is included to soften the density dependence of the symmetry energy above saturation density.
%
%
%
Our calculations will be restricted to the mean-field approximation. Hence, only the time components of the $\omega$ and $\rho$ vector meson fields are different from zero at equilibrium:
\begin{equation}
V^{(0)}_\mu=V_0 \delta_{\mu 0} ~,~ \vec{b}^{\;(0)}_\mu = b^{(0)}_\mu\; \delta_{3 i}
~,~ b^{(0)}_\mu=b_0 \delta_{\mu 0}  \ . 
\end{equation}
For the vector potential, we adopt the Landau gauge: $A^{(0)}_\mu= B x_2 \delta_{\mu 3}$, resulting in the external strong magnetic field: $\bm{B}=B\hat{e}_3$ with $\nabla \cdot \bm{B}$ = 0. The other vector fields $\omega^\mu$ and $ \rho^\mu$ are massive mesons and satisfy $\partial^\mu X_\mu=0$. The magnetic field is treated as an external, static field and is not perturbed by the proton motion. The magnetization of the matter is very small as shown in Ref.~\cite{Broderick2000}.  We designate 'strong' magnetic fields as those that correspond to a few Landau levels being occupied. 'Moderate' fields correspond to many levels being filled. 
%
%

From the Euler-Lagrange equations, we obtain the Dirac equation for the fermion fields:
\begin{eqnarray}
 i \gamma^\mu D^{(j)}_{\mu}~\psi^{(j)} = M^{\star}_{j} ~\psi^{(j)} \ , \label{dirac1}
\end{eqnarray}
and its conjugate equation:
\begin{eqnarray}
 \bar{\psi}^{(j)} i D^{\dagger (j)}_{\mu}\gamma^\mu ~= - M^{\star}_{j} ~\bar{\psi}^{(j)} \, 
 \label{dirac2}
\end{eqnarray}
where $ iD^{\dagger (j)}_{\mu} = ~ i\overleftarrow{\partial}_{\mu} + {\cal V}^{(j)}_{\mu} $. The operator with an arrow pointing to the left indicates that the derivative acts on the function to its left, in contrast to the usual right-acting derivative, i.e.,
$$ A \, \overleftarrow{\partial}_{\mu} \equiv \frac{\partial A}{\partial x^{\mu}}, \qquad \overrightarrow{\partial}_{\mu} A \equiv \frac{\partial A}{\partial x^{\mu}}$$. 

In the following section, we extend our investigation of npe matter by deriving the Vlasov equation for a hadronic system in a strongly magnetized medium, starting from the general transport equations. Our formalism employs the covariant Wigner function under strong magnetic fields, following its development in quantum electrodynamics~\cite{Vasak,Heinz}, relativistic heavy-ion physics~\cite{Mosel1,Mosel2}, and quantum chromodynamics~\cite{Elze}, as well as its application to relativistic electron gases~\cite{Hakim1,Hakim2,Hakim3,Hakim4}. For brevity, we present only the main results relevant to transport theory.
%
%
%
%
%
\subsection{Covariant Vlasov approach}
Next, we outline the key technical aspects involved in applying the covariant Wigner function formalism to describe magnetized asymmetric nuclear matter. 
A detailed derivation of the Vlasov formalism may be found in \cite{avancini2018} and references therein.

We demonstrate that the longitudinal mode, corresponding to small oscillations along the direction of the external magnetic field, can be systematically derived by extending techniques from magnetized plasma physics. The present framework is adequate for the analysis of collective modes and provides a foundation for investigating transport properties such as thermal and electrical conductivity in {\it npe} matter.
An appropriate starting point to obtain the transport equation for the (npe) matter is the Wigner covariant matrix operator \cite{Heinz} defined as:
\begin{equation}
 \hat{W}^{(j)}_4(x,p) = \int d^4y ~e^{-i p \cdot y} ~ \Phi^{(j)}_4 (x,y) \ , \label{wigner}
\end{equation}
where $x$ and $y$ are 4-vectors, $j=(n,p,e)$, and 
\begin{eqnarray}
 && \Phi^{(j)}_4 (x,y)_{\alpha \beta} \equiv 
 \bar{\psi}^{(j)}_\beta(x) e^{ \frac{y}{2} \cdot D^{\dagger (j)} } 
   e^{ - \frac{y}{2} \cdot D^{(j)} } \psi^{(j)}_\alpha (x)     \nonumber \\  
  && = \bar{\psi}^{(j)}_\beta(x_+) e^{-i y \cdot \int_{-1/2}^{1/2}~ds~ {\cal V}^{(j)}(x+ys)  }
  ~\psi^{(j)}_\alpha (x_-) \ .   \label{phi0}
\end{eqnarray}  
In the latter equation, $x_\pm=x \pm \frac{1}{2}y$, and the phase factor contains a line integral that is to be calculated along a straight line path from $x_-$  to $x_+$, since in this case the gauge invariance of the Wigner function is guaranteed \cite{Vasak}. If one defines the canonical, $\hat{p}_\mu$ and kinetic, $\hat{\Pi}_\mu$, momentum operators as:
\begin{equation}
  \hat{p}_\mu=\frac{1}{2i}(\overleftarrow{\partial}_{\mu} -\partial_\mu ) ~~, ~~
  \hat{\Pi}_\mu = \frac{1}{2i}( D^{\dagger (j)}_\mu - D^{(j)}_\mu ) = \hat{p}_\mu - 
  {\cal V}^{(j)}_\mu \ ,
   \nonumber
\end{equation}
then, from Eqs.~(\ref{wigner},\ref{phi0}), the Wigner matrix is rewritten as:
\begin{equation}
 \hat{W}^{(j)}_4(x,p)_{\alpha\beta} = (2\pi)^4 \bar{\psi}^{(j)}_\beta(x) 
 \delta^4 (\hat{\Pi} (x) - p) 
 \psi^{(j)}_\alpha (x)
 \ , \nonumber
\end{equation}
where 
$$
\delta^4(\hat{\Pi} - p)
\;\equiv\;
\frac{1}{(2\pi)^4} \int d^4y\, e^{-i y_\mu (\hat{\Pi}^\mu - p^\mu)}.
$$
One may interpret $\hat{W}^{(j)}_4(x,p)_{\alpha\beta}$ as a density matrix associated with particles in $x$ with kinetic momentum $p$.   
%
%
From the Dirac equations (\ref{dirac1},\ref{dirac2}), in a rather cumbersome calculation \cite{Vasak,Heinz}, it is possible to derive the following equation for the Wigner operator:
\begin{equation}
 \left[ \gamma^\mu (\hat{\Pi}_\mu + \frac{i\hbar}{2} {\cal D}_\mu) - 
 M^\star (x - \frac{i\hbar}{2} \partial_p)   \right] \hat{W}_4^{(j)}(x,p)=0 \ , \label{wignereq1}
\end{equation}
where
\begin{eqnarray}
 {\cal D}_\mu=\partial_{x \mu} - \int_{-1/2}^{1/2}~ds~ {\cal F}_{\mu \nu} (x+ i \hbar s \partial_p) 
 \partial_p^\nu
 \ ,
\end{eqnarray}
\begin{eqnarray}
 \hat{\Pi}_\mu= p_{\mu} + i\hbar \int_{-1/2}^{1/2}~ds~ s 
   {\cal F}_{\mu \nu} (x+ i \hbar s \partial_p) \partial_p^\nu ~,
\end{eqnarray}
and $~{\cal F}_{\mu \nu}^{(j)}= \partial_{x \mu} {\cal V}_\nu^{(j)} - \partial_{x \nu} {\cal V}_\mu^{(j)}$, where the operators $\partial_{x \mu}$ and $\partial_{p}^{\mu}$ denote partial derivatives with respect to the space-time and momentum four-vectors, respectively. 
Here, $\hbar$ appears explicitly in the expressions, as the analysis in this work is carried out to order $O(\hbar)$ to derive the semiclassical Vlasov equation. From Eq.~(\ref{wignereq1}), one obtains:  
\begin{eqnarray}
&& ({p^0}^2 - {\bm p}^{~2}-M^\star(x)^2 )~ \hat{W}_4^{(j)}(x,p)=0  \; , \; O(\hbar^0) \; .
\end{eqnarray}
\begin{equation}
 \hat{W}_4^{(j)}(x,p)=  \hat{W}_{4+}^{(j)}\delta (p_0 - E_p) +
 \hat{W}_{4-}^{(j)} \delta (p_0 + E_p) \ ,  \label{onmass}
\end{equation}
where $E_p=\sqrt{{\bm p}^{~2}+ M^\star(x)^2}$. From now on, we will omit the particle index $(j)$ whenever this does not cause confusion.

For the derivation of the Vlasov equation, the Wigner $4\times 4$ matrix operator is first expanded on the basis of the Clifford algebra Ref.~\cite{Vasak}:
\begin{eqnarray}
 \hat{W}_4(x,p)&=&\frac{1}{4} \left[ F(x,p) I + i\gamma^5 P(x,p) +\gamma^\mu F_\mu(x,p) \right.
                                             \nonumber \\
&& \left. + \gamma^\mu \gamma^5 \Omega_\mu (x,p) + \frac{1}{2} \sigma^{\mu \nu} 
                           S_{\mu \nu}(x,p)
  \right]  \ ,  \label{cliff}
\end{eqnarray}
where $\sigma^{\mu \nu}= \frac{1}{2}( \gamma^\mu \gamma^\nu-\gamma^\nu \gamma^\mu)$.
After substituting the latter equation in Eq.~(\ref{wignereq1}), one obtains:
\begin{eqnarray}
&& \left[ \gamma^\mu (\hat{\Pi}_\mu + \frac{i\hbar}{2} {\cal D}_\mu) - 
 M^\star (x - \frac{i\hbar}{2} \partial_p)   \right]
 \nonumber \\
&& \times  \frac{1}{4} \left[ F(x,p) I + i\gamma^5 P(x,p) +\gamma^\mu F_\mu(x,p) \right.
                                             \nonumber \\
&& \left. + \gamma^\mu \gamma^5 \Omega_\mu (x,p) + \frac{1}{2} \sigma^{\mu \nu} 
                           S_{\mu \nu}(x,p)
  \right] = 0 \ . 
\end{eqnarray}
For convenience, we define $\hat{W}_3(x,\bm{p})$ as:
\begin{equation}
 \hat{W}_3(x,\bm{p})=\int d p_0 \hat{W}_4 (x,p^0,\bm{p})~\gamma^0 \ .
\end{equation}
From the evaluation of the Dirac traces 16 complex equations are obtained \cite{Heinz}. After appropriate manipulations, the general kinetic equation relevant for the derivation of the Vlasov equation is given by:
\begin{equation}
 D_t \; f_0 + \bm{D}\cdot \bm{g}_1 = 2 \int dp_0 \; \Im(M^\star)\; F \; , 
\end{equation}
where
\begin{eqnarray}
&& f_0(x,\bm{p}) = \int dp_0 \; F_0(x,p) = \int dp_0 \; {\text Tr} [ \gamma^0\; \hat{W}_4 ] 
\; , \nonumber \\
&& \bm{g}_1 (x,\bm{p}) = \int dp_0 \bm{F} (x,\bm{p}) = \int dp_0 {\text Tr} [\boldsymbol{\gamma} \; \hat{W}_4 ] \; ,
\end{eqnarray}
and $\Im(M^\star)$ stands for the imaginary part of $M^\star (x - \frac{i\hbar}{2} \partial_p)$.
Finally, we consider the approximation when only terms up to $O(\hbar)$ are included. After a long and tedious calculation \cite{Heinz}, the generalized Vlasov equation is obtained:
 \begin{eqnarray}
  && \partial_t f^{(j)}_{\pm} \pm \bm{v}^{(j)} \cdot \nabla_x f^{(j)}_{\pm} +
  (\bm{\mathcal{E}}^{(j)} \pm \bm{v}^{(j)} \times \bm{\mathcal{ B}}^{(j)})\cdot \nabla_p f^{(j)}_{\pm} \nonumber \\
  && \mp \frac{M^\star}{E^{(j)}_p} \nabla_x M^\star \cdot \nabla_p f^{(j)}_{\pm} = 0  \ ,
  \label{vlasov}
 \end{eqnarray}
where the quantity $E^{(j)}_p=\sqrt{\bm{p}^2+{M_j^{\star}}^2}$ represents the single-particle energy of particle $j$, so that $\bm{v}^{(j)}=\bm{p}/E^{(j)}_p$ gives the ratio between the particle’s spatial momentum and its energy, thereby describing its velocity.
The electric-like and magnetic-like components acting on each particle $j$ are defined as
\begin{eqnarray}
 && {\cal E}_i^{(j)} = {\cal F}_{0i}^{(j)} \mp \frac{M^\star(x)}{E_p^{(j)}} 
 \nabla_{x,i}~ M^\star(x) ~\ , \  j=p,n \nonumber \ , \\
 && {\cal E}_i^{(e)} = {\cal F}_{0i}^{(e)}  \nonumber \ , \\
 && {\cal B}^{(j)}_i ~=~\epsilon_{ilm} \partial_{x,l} {\cal V}^{(j)}_m ~,~j=p,n,e~\ , \label{emlf}
\end{eqnarray}
$i,l,m=1,2,3$ and particle and antiparticle distribution functions $f^{(j)}_{\pm}$ are given by
\begin{equation}
 f^{(j)}_{\pm}(x,\bm{p}) 
 =\frac{1}{(2\pi)^4} \int dp_0 \; Tr[ \hat{\rho} \gamma^0 
 \hat{W}^{(\pm)}_4] \ ´ \nonumber \\
\end{equation}
with $\hat W_4^\pm$ defined in Eq. (\ref{onmass}).
At finite-temperature, the signs $\pm$ appearing in Eq.~(\ref{vlasov}) and Eq.~(\ref{emlf}) are assigned as follows: the top (bottom) sign is for particles (antiparticles).

Next, we use the density matrix operator,
\begin{equation}
 \hat{\rho}= \frac{1}{Z} e^{-\beta (\hat{H}-\sum_j \mu_j \hat{N}_j )} \ ,
\end{equation}
where $Z$ is the partition function, $\beta=1/T$, and the trace is to be evaluated simultaneously in the Dirac space and in any arbitrary set of basis states, for example, 
Johnson-Lippmann \cite{john}. 
The relativistic Hamiltonian for npe matter at equilibrium in the mean-field approximation is given by 
$
\hat{H}= \hat{H}_{fermion}+\hat{H}_{meson}+ \hat{H}_{\cal A},
$
with
\begin{eqnarray}
\hat{H}_{fermion} &=& \sum_{j=p,n,e} \int d^{3}x\; \hat{\psi}_{j}^{\dagger}(x) \left[-i\boldsymbol{\alpha}\cdot \left(\nabla - i\; \boldsymbol{\mathcal{V}}^{(0)(j)} \right) \right. \nonumber\\
&+& \left.\beta M_j^{*}(x) + \mathcal{V}^{(0)(j)}_{0}(x) \right] \hat{\psi}_{j}(x).  
\end{eqnarray}
where, $\boldsymbol{\alpha}$, $\beta$ are the Dirac matrices, and $\hat{H}_{\rm meson}$ and $\hat{H}_{\cal A}$ are the mesonic and electromagnetic contributions to the Hamiltonian, defined as usual from the Lagrangian density, $$
\hat{H}_{meson, A} =\int d^3 x \left(\sum_i\frac{\partial \mathcal{L}'}{\partial(\partial_0 F_i)} \dot{F}_i -\mathcal{L}'\right),  $$
with $F$ representing the scalar and vector fields of the theory $F=\sigma, \omega^\mu, \vec b^\mu, A^\mu,$ and $\mathcal{L}'= \cal L_\sigma + {\cal L}_\omega +
{\cal L}_\rho + {\cal L}_{\omega \rho } + {\cal L}_{A}$
contains  the non-fermionic terms of the Lagrangian density given by Eq.~(\ref{lag}).

We obtain the equilibrium Wigner distribution function as follows:
\begin{equation}
 f_{\pm}^{(0)(j)}(x,\bm{p}) =\frac{1}{(2\pi)^4} \int dp_0 Tr[ \hat{\rho}~ \gamma^0 
 \hat{W}^{(\pm)(0)}_4] \ , \nonumber \\
\end{equation}
The main advantage of the present method is that we have a systematic way to obtain the Wigner distribution function. In particular, when an external magnetic field is present, this issue is crucial. 
The Vlasov equation has previously been derived in a Hamiltonian formalism~\cite{nielsen89,nielsen91}, however, how to obtain the distribution in the presence of external fields is not included in the formalism.   
Of course, $f^{(0)(j)}_{\pm}(x,\bm{p})$ satisfies the Vlasov equation by its construction.
Substituting the explicit form of the Wigner operator, Eq.~(\ref{wigner}), and the on-mass-shell constraint, Eq.~(\ref{onmass}), one obtains:
\begin{eqnarray}
 &&f^{(0)(j)}_{\pm}(x,\bm{p}) =\int \frac{d p_0}{(2\pi)^4} Tr[ \hat{\rho}~ \gamma^0
                             \int d^4 y      e^{-i p\cdot y}\nonumber \\ 
 &&\bar{\psi}^{(j)}_{\pm}(x_+) e^{-i y \cdot \int_{-1/2}^{1/2}~ds~ {\cal V}^{(j)}(x+ys)  }
  ~\psi^{(j)}_{\pm} (x_-)    ] \ ,  \label{wignereq}
\end{eqnarray}
where $\psi^{(j)}_{\pm}(x)$, $j=(n,p,e)$, are the positive and negative components of the
fermionic Dirac fields associated with the neutron, proton, and electron.
We restrict our calculation to the mean-field approximation, then the integral in the phase factor, Eq.~(\ref{wignereq}), results in ${\cal V}^{(0)(j)}_{\mu}$, therefore:
\begin{eqnarray}
 &&f_{\pm}^{(0)(j)}(x,\bm{p}) =\int \frac{d p_0}{(2\pi)^4} \int d^4 y  
            Tr[ \hat{\rho}~ \gamma^0
                                 e^{-i (p+{\cal V}^{(0)(j)}) \cdot y}\nonumber \\ 
 &&\bar{\psi}^{(j)}_{\pm}(x_+) 
  ~\psi^{(j)}_{\pm} (x_-)    ] \ . \nonumber \\ \label{wignereq3}
\end{eqnarray}
The equilibrium Wigner function, as given in the latter equation, may be calculated using standard techniques, resulting in the following expression:
\begin{eqnarray}
 &&f_{\pm}^{(0)(j)}(\bm{p}) = \frac{2}{(2\pi)^3} \;\sum_{n=0}^\infty  
            \left[ L_{n}(2w^2)-L_{n-1}(2w^2) \right] \nonumber \\ 
 &&\times \frac{ (-1)^{n} ~e^{-w^2}}{1+\exp\left[ \beta(E_n \mp \tilde{\mu}^{(j)})\right] }
 \ , j=p,e \nonumber \\ 
 \label{wignereq4}
\end{eqnarray}
where $w^2=\frac{p_\perp^2}{eB}=\frac{p_1^2+p_2^2}{eB}$, $L_n(x)$ are the Laguerre polynomials, $\tilde{\mu}^{(e)} = \mu_e $ and $\tilde{\mu}^{(p)} = \mu_p - {\cal V}^{(0)(p)}_0 $ are the effective chemical potential of the electron and proton, respectively, $ E_n=\sqrt{p_3^2+{M^{\star}_j }^2 + 2eBn }$ where $n$ is the Landau level label, and we have defined $L_{-1}(x)$=0.

In plasma physics, a similar calculation has been derived for a magnetized electron gas~\cite{Hakim1}. Since in the present work we are interested only in systems at zero temperature ($T=0$), the previous expression can be written in the zero-temperature limit as:
\begin{eqnarray}
 &&f^{(0)(j)}(\bm{p}) = \sum_{n=0}^\infty  
            \left[ L_{n}(2w^2)-L_{n-1}(2w^2) \right]  \nonumber \\ 
 &&\times (-1)^{n}~ e^{-w^2} \theta( \tilde{\mu}^{(j)} -E_n)
 \ ,   \label{wignereq5}
\end{eqnarray}
where $\theta(x)$ is the Heaviside step function. For convenience of notation, the factor $\frac{2}{(2\pi)^3}$ is omitted from the definition of $f^{(0)(j)}(\bm{p})$; this factor will be incorporated into the integration over momenta in what follows.
The limit $T \to 0$ rules out the negative energy part of the distribution function, and therefore we will omit the redundant plus signal in the symbol $f_{+}^{(0)(j)}$.
The electron or proton equilibrium densities can readily be calculated from their corresponding distribution functions. For example, the electron density reads from Eq.~(\ref{wignereq5}):
\begin{eqnarray}
 &&\rho^{(0)}_e=\frac{2}{(2\pi)^3} \int d^3 p~ f^{(0)(e)}(\bm{p})
 \nonumber
\\  
&& = \frac{eB}{2\pi^2} \sum_{n=0}^{n_{max}} g_n~ p_F^{(e)}(n) \label{densie} \; .
\end{eqnarray} 
with $n_{max}=[\frac{\mu_e^2-m_e^2}{2eB}]$ where [...] stands for the floor function, which gives the largest integer that is less than or equal to x. $g_n$ is a degeneracy factor that is 1 for $n=0$ and 2 for $n \geq 0$, and $p_F^{(e)}(n)=\sqrt{\mu^2_e - m_e^2-2eBn}$.
The density of electrons just obtained coincides with the usual one \cite{broderick}.
The neutron distribution function is also the standard one:
$f^{(0)(n)}(\bm{p}) = \theta(p_F^{(n)~2}-\bm{p}^{~2})$. 
\subsection{Dispersion relations}
%
%
%
%
The conserved fermion 4-current, at zero-temperature, is defined  as: 
\begin{equation}
  J^{(j)}_\mu (x) = \frac{4}{(2\pi)^3} \int d^4p\; p_\mu \; f^{(j)}(x,p) \; .
\end{equation}
Next, we use the on mass shell constraint of the distribution function: 
\begin{equation}
  f^{(j)}(x,p)=\theta (p^0)\; \delta \left(p^2 - M^{\star \;2}_j \right) \;f^{(j)} (x,\bm{p}) \; . 
\end{equation}
In the latter equation, the theta function is used to restrict ourselves to the case of the positive energy part of the distribution function, which is the scope of our work. The integral over the variable $p^0$ may be performed using the previous equation, resulting in:
\begin{equation}
 J^{(j)}_\mu(x)=\frac{2}{(2\pi)^3} \int \frac{d^3 p}{p^0}\; p_\mu 
 f^{(j)}(x,\bm{p}) \label{curr} \ ,
\end{equation}
where $j=(n,p,e)$, and  {$p^0=E_j=\sqrt{\bm{p}^{~2}+M^{\star~2}_j}$}, and the factor 2 accounts for spin degeneracy of protons, neutrons and electrons. The state with the quantum number $p_\mu$ represents a single-particle momentum state.

From the Vlasov equation, Eq.~(\ref{vlasov}), the following 4-current conservation law follows: $\partial^\mu  J^{(j)}_\mu(x)$ = 0. In our formalism, $p^\mu$ represents the four-momentum of the particle, where the temporal component corresponds to the single-particle energy and the spatial components to the particle momentum. The quantity 
$J^\mu$ denotes the particle four-current density, whose time component gives the particle (number) density and whose spatial components correspond to the particle current density.

The dispersion relations are derived from the Vlasov equation by considering a small perturbation around the equilibrium distribution functions as follows:
\begin{equation}
  f^{(j)}(\bm{x},\bm{p},t) = f^{(0)(j)}(\bm{p}) + \delta f^{(j)} \,
  \label{delphi},
\end{equation}
which in turn generates a corresponding perturbation of the equilibrium 4-current,
\begin{equation}
  J^{(j)}_\mu(x) = J^{(0)(j)}_\mu
   + \delta J^{(j)}_\mu \ ,
\end{equation}
with
\begin{eqnarray}
&& J^{(0)(j)}_\mu = \frac{2}{(2\pi)^3} \int \frac{d^3 p}{p^0} p_\mu 
f^{(0)(j)}(\bm{p})   \ , \nonumber
\\ 
&&  \delta J^{(j)}_\mu = \frac{2}{(2\pi)^3} \int \frac{d^3 p}{E^{(0)}_j}
    ~p_\mu ~\delta f^{(j)}   \ , \label{curre}
\end{eqnarray}
where we have used the notation $p^0= E^{(0)}_j=\sqrt{\bm{p}^{~2}+{M_j^{\star (0)}}^2}$, $j=(n,p,e)$, with $M_p^{\star (0)}$ = $M_n^{\star (0)}$ = $M-g_s \phi^{(0)}$ and $M_e^{\star (0)}$=$m_e$. 
In addition, a small perturbation of the distribution functions, $f^{(0)(j)}$, around their equilibrium values, will generate perturbations in the fields:
\begin{eqnarray}
&& \phi = \phi^{(0)}+\delta \phi ~,~V_\mu=V_\mu^{(0)}+\delta V_\mu ~,~
 b_\mu=b_\mu^{(0)}+\delta b_\mu~, \nonumber \\
 &&~A_\mu=A_\mu^{(0)}+ \delta A_{\mu}~ . \label{smalldev}
\end{eqnarray}
%
%
%
The scalar density is given by the expression:
 \begin{equation}
  \rho_s^{(j)}= \frac{2}{(2\pi)^3} \int \frac{d^3 p}{E_j} {M^\star}
 f^{(j)}(x,\bm{p})    \ .
 \end{equation}
The small perturbation of the distribution function induces for the scalar density:
\begin{equation}
\rho^{(j)}_s(x) = \rho^{(0)(j)}_s(x) + \delta \rho^{(j)}_s \; .
\end{equation} 
with
\begin{equation}
  \rho_s^{(0)(j)}= \frac{2}{(2\pi)^3} \int \frac{d^3 p}{E^{(0)}_j} {M^\star}^{(0)} 
 f^{(0)(j)}(\bm{p})    \ , 
\end{equation}
For the scalar density, the calculation is more involved~\cite{avancini05}, since ${M^\star}= M-g_s \phi(x)$ is position dependent, resulting in the following expression:
\begin{equation}
 \delta \rho_s^{(j)}= \delta\tilde{\rho}_s^{(j)}+ g_s ~d\rho_s^{(0)(j)}~ \delta \phi\ ,
\end{equation}
with:
 \begin{equation}
  \delta \tilde{\rho}_s^{(j)}= \frac{2}{(2\pi)^3} \int \frac{d^3 p}{E^{(0)}_j} 
  {M^\star}^{(0)} \delta f^{(j)}    \ ,  \label{densca}
 \end{equation}
 and
 \begin{equation}
  d\rho_s^{(0)(j)} = \left(\frac{\partial \rho^{(j)}_s}{\partial M^*}\right)_{(0)}=- \frac{2}{(2\pi)^3} \int d^3 p 
  \frac{{\bm{p}}^2} {{E^{(0)}_j}^3}   f^{(0)(j)}    ~.
 \end{equation}
\noindent
To obtain the dispersion relation, we substitute Eq.~(\ref{delphi}) in the Vlasov equation, Eq.~(\ref{vlasov}) and retain only terms of the first order in $\delta f^{(j)}$:
 \begin{eqnarray}
  && \partial_t \delta f^{(j)} + \bm{v}^{(j)} \cdot \nabla_x \delta f^{(j)} +
  \bm{v}^{(j)} \times ( \nabla_x \times \bm{\mathcal{V}}^{(0)(j)} ) \cdot \nabla_p \delta f^{(j)}
  \nonumber \\
  && + \left[ \bm{v}^{(j)} \times \nabla_x \times ({\bm{\mathcal{V}}}^{(0)(j)}  + 
\delta {\bm{\mathcal{V}}}^{(j)} ) + g_s \frac{{M^\star}^{(0)}}{{E^{(0)}_j}} \nabla_x\delta \phi 
\right. \nonumber \\
 && \left.  
 \nabla_x \delta {\cal V}_0^{(j)}  \right]  \cdot 
 \nabla_p f^{(0)(j)}    
   = 0 \ ,
  \label{dvlasov}
 \end{eqnarray}
where $\bm{v}^{(j)}=\bm{p}/E^{(0)}_j$, $j=p,e$ (for electrons $g_s=0$). 
%
%
First, we consider the Fourier transform of the small deviation from equilibrium of the distribution function and fields:
 \begin{equation}
 \left\{
       \begin{array}{c}
          \delta f^{(j)} (\bm{x},\bm{p},t) \\
         \delta \phi (\bm{x},t) \\
         \delta {\cal V}^{(j)}_{\mu} (\bm{x},t)
       \end{array}    
 \right\}  = 
 \int d^3q ~d\omega 
  \left\{
       \begin{array}{c}
          \delta f^{(j)} (\bm{q},\omega,\bm{p}) \\
          \delta \phi (\bm{q},\omega) \\
          \delta {\cal V}^{(j)}_{\mu}(\bm{q},\omega) 
       \end{array}    
 \right\}  e^{i(\omega t - \bm{q} \cdot \bm{x})} \ ,
 \end{equation}
The dispersion relations are obtained by substituting the latter equations in the Vlasov equation, Eq.(\ref{dvlasov}), one obtains for $ \delta f^{(j)}(\bm{q},\omega,\bm{p})$ :
 \begin{eqnarray}
  && i (\omega - \bm{v}^{(j)} \cdot \bm{q} ) ~\delta f^{(j)} - \frac{Q_j B}{E^{(0)}_j}
  \frac{\partial}{\partial \Phi}  \delta f^{(j)} \nonumber \\
  && = i \left[ (\omega - \bm{v}^{(j)} \cdot \bm{q} ) ~\delta \bm{\mathcal{ V}}^{(j)}  -
  \left( \delta {\cal V}_0^{(j)} - \bm{v}^{(j)} \cdot \delta \bm{{\cal V}}^{(j)}
   \right.   \right. \nonumber \\
 &&  \left. - g_s \frac{{M^\star}^{(0)}}{{E^{(0)}_j}} \delta \phi \right) \bm{q}\left.   
 \right]  \cdot 
 \nabla_p f^{(0)(j)},
  \label{dvlasov3}
 \end{eqnarray}
where $Q_j$ is the electrical charge of species $j$ and $\Phi$ is the angle in the spherical coordinates.
%
%
%

%
In the latter equation, an explicit dependence on the angle $\Phi$ is present. Here, we can use some techniques used in magnetized plasma physics \cite{Kelly,Oberman} to get rid of this dependence. Adopting the same reference frame as \cite{Kelly}
\begin{eqnarray}
 && \bm{B}=B \bm{e}_3 ~,~\hat{e}_3 \equiv e_\parallel  \nonumber \\
 && \bm{q}= q_\perp \bm{e}_1 + q_\parallel \bm{e}_3 = (q_\perp,0,q_\parallel) \nonumber \\
 && \bm{p}= p_\perp \bm{e}_\perp + p_\parallel \bm{e}_3 = 
 (p_\perp \cos \Phi,p_\perp \sin \Phi, p_\parallel )  ~ .
\end{eqnarray}
we can use the Oberman-Ron \cite{Oberman} transform:
\begin{eqnarray}
 &&\delta f^{(j)} (\bm{q},\omega,\bm{p}) = \nonumber \\
 && e^{i b \sin \Phi} \sum_{m=-\infty}^{\infty} e^{-im\Phi} J_m(b) 
 \delta f^{(j)}_m (\bm{q},\omega,p_\parallel,p_\perp) \ , \label{trans} 
\end{eqnarray}
and its inverse transform:
\begin{eqnarray}
 &&\delta f^{(j)}_m (\bm{q},\omega,p_\parallel,p_\perp) = \nonumber \\
 &&\frac{1}{2\pi J_m(b)} \int_0^{2\pi} d\Phi e^{-i b \sin \Phi} e^{im\Phi} 
 \delta f^{(j)} (\bm{q},\omega,\bm{p}) ~ , \label{invtrans} 
\end{eqnarray}
where $J_m(b)$ is a Bessel function and $b$ is an arbitrary real constant. 
We substitute the 
Oberman-Ron transform, Eq.~(\ref{trans}), in the Vlasov equation, Eq.~(\ref{dvlasov3}),
with $b=-\frac{q_\perp p_\perp}{Q_j B}$, $j=(p,e)$, and integrate in $\Phi$ the resulting
expression multiplied by the factor $\exp[-i(b\sin \Phi-m\Phi)]$. After some straightforward
but tedious calculations, we finally obtain, using Eq.~(\ref{invtrans}), the following expression:
\begin{eqnarray}
&&  \delta f^{(j)} (\bm{q},\omega,\bm{p})= e^{i b \sin \Phi} \sum_{m=-\infty}^{\infty}
  e^{-im\Phi}   \nonumber \\
&&\times \left\{  \frac{m}{b} J_m(b)\left[\omega D_\perp^{(j)}- \frac{q_\parallel}{E^{(0)}_j}
(p_\parallel D_\perp^{(j)} -p_\perp D_\parallel^{(j)}) \right]   \delta {\cal V}^{(j)}_x \right.  \nonumber \\ 
&&  + i J_m^\prime (b)\left[\omega D_\perp^{(j)}- \frac{q_\parallel}{E^{(0)}_j}
(p_\parallel D_\perp^{(j)} -p_\perp  D_\parallel^{(j)}) \right]
                                                                 \delta {\cal V}^{(j)}_y   \nonumber \\   
&&  + J_m (b)\left[\omega D_\parallel^{(j)}- \frac{q_\perp }{E^{(0)}_j}
(p_\perp D_\parallel^{(j)} -p_\parallel  D_\perp^{(j)}) \frac{m}{b} \right]
                                                                 \delta {\cal V}^{(j)}_z   \nonumber \\ 
&& \left. - J_m (b)\left[ q_\perp D_\perp^{(j)} \frac{m}{b} + q_\parallel D_\parallel^{(j)} \right] 
\left[ \delta {\cal V}^{(j)}_0 -\frac{g_s M^{\star(0)}}{E^{(0)}_j} \delta \phi \right] \right\}
\nonumber \\    
&& \times \left[ \omega - \frac{p_\parallel q_\parallel}{E^{(0)}_j} + \frac{Q_j B m}{E^{(0)}_j} 
\right]^{-1}
\label{dispvla}
\end{eqnarray}
where the prime of the Bessel function means its derivative with respect to $b$, and
\begin{equation}
 D_\perp^{(j)} = \frac{\partial}{\partial p_\perp} f^{(0)(j)}~,~ 
     D_\parallel^{(j)} = \frac{\partial}{\partial p_\parallel} f^{(0)(j)} ~ .
\end{equation}
The expression above is the key for the calculation of the dispersion relations in the presence of a strong external magnetic field. The Fourier transform of the small deviations from equilibrium of the current is given by:
\begin{equation}
   \delta J^{(j)}_\mu (\bm{x},t)=   \int d^3q ~d\omega
        ~  \delta J^{(j)}_\mu (\bm{q},\omega) 
        e^{i(\omega t - \bm{q} \cdot \bm{x})} \ .
\end{equation}
From Eq.~(\ref{curre}) it follows that:
\begin{equation}
   \delta J^{(j)}_\mu (\bm{q},\omega) = \frac{2}{(2\pi)^3} \int \frac{d^3 p}{E^{(0)}_j}
    ~p_\mu ~\delta f^{(j)}  (\bm{q},\omega,\bm{p})  \ . \label{delcurr}
\end{equation}

From the current conservation immediately follows the relation:
\begin{equation}
 \partial^\mu J^{(j)}_\mu = 0 \Rightarrow \omega \delta J^{(j)}_0 + \bm{q} \cdot \delta 
\bm{J}^{(j)} =0 \ .
\end{equation}
This latter equation may also be used to obtain the very useful relation:
\begin{equation}
 \partial^\mu \delta {\cal V}^{(j)}_\mu = 0 \Rightarrow \omega \delta {\cal V}^{(j)}_0 + 
 \bm{q} \cdot \delta 
\bm{\mathcal{ V}}^{(j)} =0 \ . \label{Vcons}
\end{equation}
\subsection{Dispersion relations for longitudinal mode}
\noindent The particular case of longitudinal modes consists of small perturbations parallel to the external magnetic field. It is obtained by imposing in Eq.~(\ref{dispvla}) the following restrictions:
\begin{equation}
q_\parallel=q \ne 0 \; , \; q_\perp =0 \; , \, b=0 \; , \; \delta {\cal V}_x = \delta {\cal V}_y =0 \; .
\end{equation}
The conservation law, Eq.~(\ref{Vcons}), 
for the longitudinal modes may be written as: 
\begin{equation}
 \omega \delta {\cal V}^{(j)}_0=\bm{q} \cdot \delta \bm{\mathcal{ V}}^{(j)} = q \delta {\cal V}^{(j)}_z \ .  \label{simplify_omega}
\end{equation}
The latter results show that $\delta {\cal  V}^{(j)}_z$ can be written in terms of $\delta {\cal V}^{(j)}_0$, which considerably simplifies the final expressions.
The dispersion relation for density perturbations follows from taking $\mu=0$ in Eq.~(\ref{delcurr}):
{\small{
\begin{eqnarray}
 && \delta \rho^{(j)} \equiv \delta J^{(j)}_0 (\bm{q},\omega)
 = \frac{2}{(2\pi)^3} \int d^3 p ~\delta f^{(j)}  (\bm{q},\omega,\bm{p})  \nonumber \\
 && = \frac{2}{(2\pi)^3} \int_0^\infty dp_\perp p_\perp \int_{-\infty}^\infty dp_\parallel 
 \int_0^{2\pi} d\Phi ~\delta f^{(j)}  (\bm{q},\omega,\bm{p}) \ ,
\end{eqnarray} }}
Next, we substitute in the latter equation $\delta f^{(j)}  (\bm{q},\omega,\bm{p})$ given in Eq.~(\ref{dispvla}) and after performing the
integration in $\Phi$ and using Eq.~(\ref{simplify_omega}), one obtains:
{\small{
\begin{eqnarray}
&&  \delta \rho^{(j)} (\bm{q},\omega)=   
  -q\left(1-\frac{\omega^2}{q^2}\right) S_j \left[J_m^2 (0) D_\parallel^{(j)}  \right]  
                                  \delta {\cal V}^{(j)}_0    \nonumber \\
&&  + q S_j \left[ \frac{J_m^2 (0)}{E^{(0)}_j}  
  D_\parallel^{(j)} \right]  g_s {M^\star}^{(0)} ~ \delta \phi  
\label{displongi}  \ .
\end{eqnarray} }}
where we have defined, as in Ref. \cite{Kelly}, the function:
{\small{
\begin{equation}
 S_j \left[ X \right] = \sum_{m=-\infty}^{\infty} \frac{1}{2\pi^2}\int_{-\infty}^{\infty}
 dp_{\parallel} \int_{0}^{\infty}
 \frac{dp_\perp~p_\perp   \left[  X \right]}
      {\omega -\frac{p_\parallel q_\parallel}{E^{(0)}_j} + \frac{Q_j Bm}{E^{(0)}_j} }~,
      \label{Lind}
\end{equation} }}
where $X$ is an arbitrary function of $p_\parallel$ and $p_\perp$. 
From Eq.~(\ref{densca}), it follows that the scalar density also may be written as a function of the distribution function, analogously to the particle density fluctuations:
{\small{
\begin{equation}
\delta {\tilde \rho}^{(j)}_s 
 = \frac{2}{(2\pi)^3} \int d^3 p \frac{{M^\star}^{(0)} }{E^{(0)}_j}~
                                   \delta f^{(j)}  (\bm{q},\omega,\bm{p}) \ .
\end{equation} }}
The latter expression allows us to write the scalar density in a completely analogous way to the particle density, yielding:
{\small{
\begin{eqnarray}
&&  \delta \tilde{\rho}^{(j)}_s (q,\omega)=   
  -q\left(1-\frac{\omega^2}{q^2}\right) S_j \left[\frac{J_m^2 (0)}{E^{(0)}_j} D_\parallel^{(j)}  \right]  
                               {M^\star}^{(0)}   \delta {\cal V}^{(j)}_0    \nonumber \\
&&  + q S_j \left[ \frac{J_m^2 (0)}{{E^{(0)}_j}^2}  
  D_\parallel^{(j)} \right]  g_s {{M^\star}^{(0)}}^2 ~ \delta \phi  
\label{displongisca}  \ .
\end{eqnarray} }}
To obtain the dispersion relations, the final step is to employ  the equations of motion of the mesonic and electromagnetic fields , given in appendix A, in order to rewrite Eqs.~(\ref{displongi}) and (\ref{displongisca}) only in terms of the neutron and proton densities, their corresponding scalar densities, and the electron density as detailed in Appendix B.
%
One can show that the dispersion relations may be associated with a five-dimensional linear system, involving the 
number densities, $\delta \rho^{(p)}$, $\delta \rho^{(n)}$, $\delta \rho^{(e)}$, and the 
scalar densities $\delta \tilde{\rho}^{(p)}_s$, $\delta \tilde{\rho}^{(n)}_s$,  which may be written in matrix form as follows:
{\small{
\begin{equation}
\left(\begin{array}{ccccc}
a_{11}&a_{12}&a_{13}&a_{14}&a_{15}\\
a_{21}&a_{22}& 0 &a_{24}&a_{25}\\
a_{31}& 0 &a_{33}& 0 &0\\
a_{41}&a_{42}&a_{43}&a_{44}&a_{45}\\
a_{51} &a_{52}&0& a_{54} &a_{55}\\
\end{array}\right)
\left(\begin{array}{c}
  \delta \rho^{(p)} \\    \delta \rho^{(n)} \\
     \delta \rho^{(e)} \\ \delta \tilde{\rho}^{(p)}_s \\ \delta \tilde{\rho}^{(n)}_s
\end{array}\right)
=0 \ .
\label{det}
\end{equation} }}
The normal modes $\omega$ of matter are obtained numerically from the determinant of the matrix $\{a_{ij}\}$.  The matrix elements depend on the interactions and generalized Lindhard functions, which are nonlinear in $\omega$ and the momentum $q$. Since,  we are interested in studying the stable modes, we  will only consider the real roots of the determinant.
The entries of the matrix  $\{a_{ij}\}$ are discussed in  detail in the appendix.
%
%
\section{Results and discussion}
In the present section, we discuss the effects of strong magnetic fields on the propagation of collective nuclear longitudinal modes obtained from the solution of the dispersion relation given in Eq.~(\ref{det}). To carry out our study, we proceed in a stepwise manner. First, we focus on proton–neutron (pn) matter, where two distinct configurations are analyzed: (i) pn matter without Coulomb interaction, to isolate the contribution of the nuclear interaction; and (ii) pn matter with Coulomb interaction, to assess the role of electromagnetic coupling. Subsequently, we include the electrons and consider neutron-proton-electron (npe) matter with a fixed proton fraction $y_p=0.1$ and momentum transfer $q=10$ MeV. This sequence of steps allows us to disentangle the effects of each interaction channel on the collective response of the magnetized neutron star matter.

\begin{figure}[htbp]
\includegraphics[width=1\linewidth,angle=0]{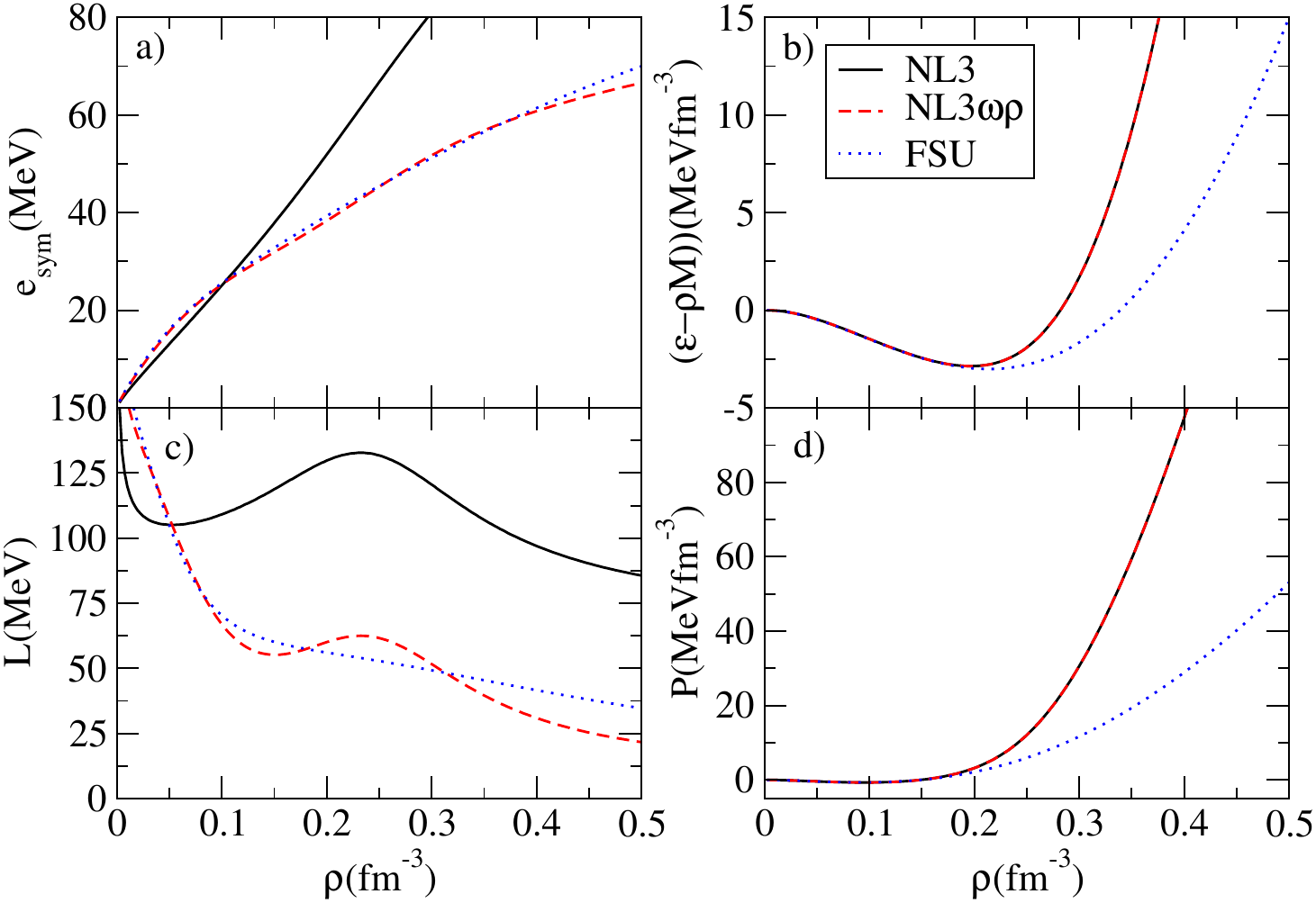}
\caption{Symmetry energy and energy density (top panels), symmetry energy slope and beta-equilibrium pressure.}
\label{ener_sym}
\end{figure}

All calculations are carried out at zero temperature, without taking into account the anomalous magnetic moments (AMM) of nucleons. We will mostly consider magnetic-field intensities $B=5\times10^{17}$G and $B=1\times10^{18}$G, although we will also show some results for $B=1\times 10^{17}$G. The most intense fields detected on the surface of a magnetar are not larger than $2\times10^{15}$G. However, it has been shown that convective dynamo convection inside a proto-neutron star can amplify the magnetic field, giving rise to turbulent magnetic fields with an average strength that can reach $\sim 10^{16}$~G \cite{Masada:2020dpy,Reboul-Salze:2020mnw}.
Moreover, toroidal fields more intense than $10^{17}$G  were obtained in stable configurations ~\cite{kiuchi08, rezzolla12}, and, therefore, stronger fields may be expected in the interior of the stars.

\begin{figure*}[t]
\begin{tabular}{cc}
\includegraphics[width=0.5\linewidth,angle=0]{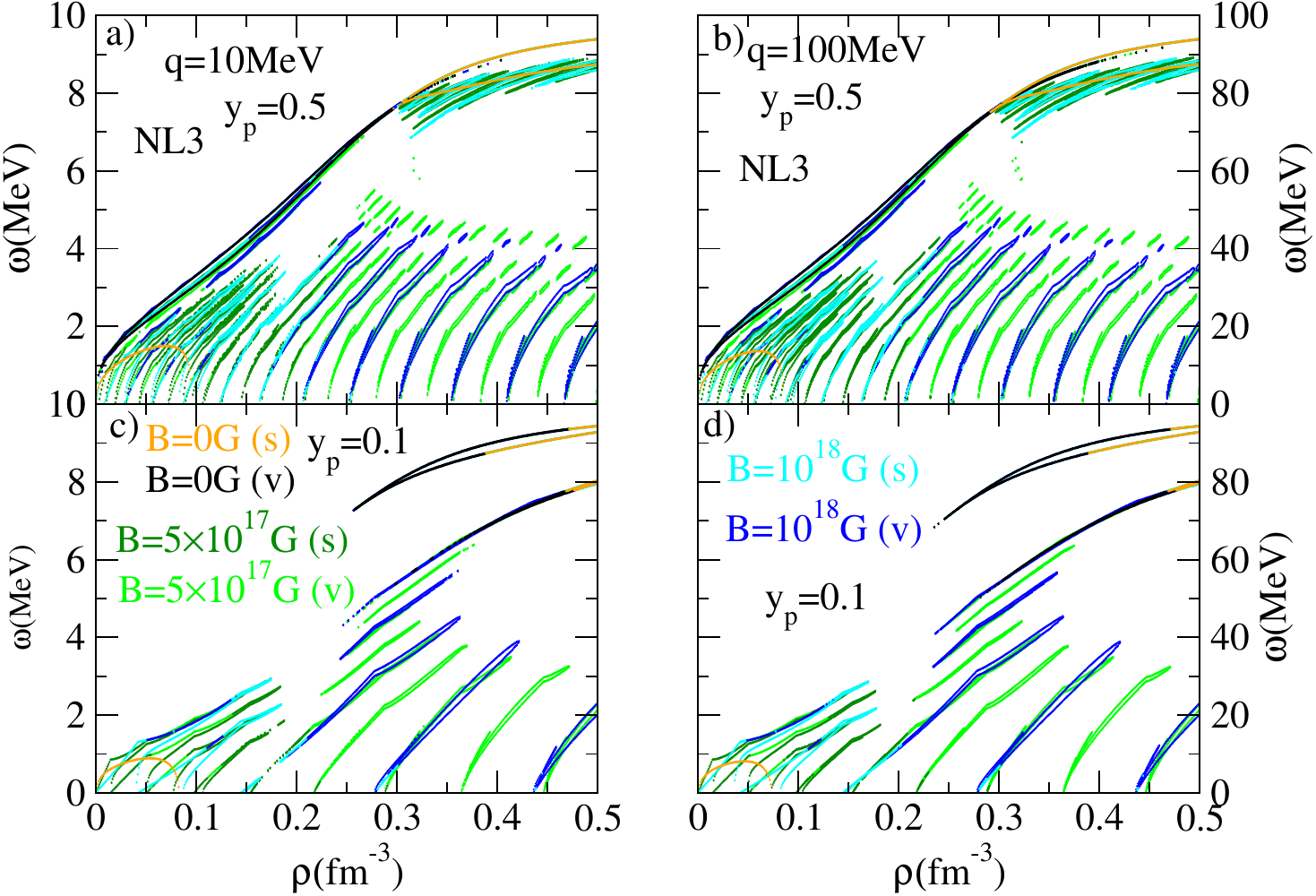}&
\includegraphics[width=0.5\linewidth,angle=0]{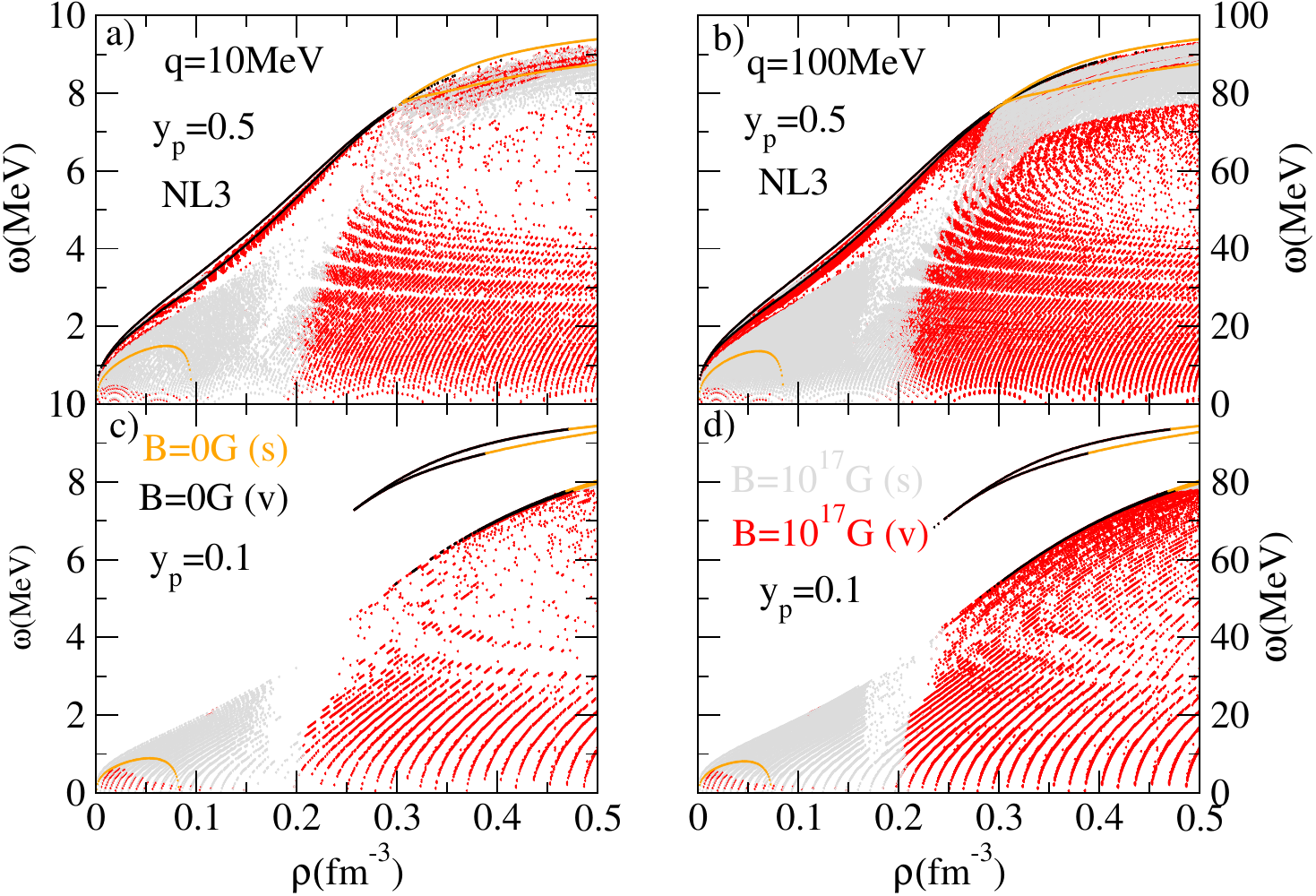}\\
\includegraphics[width=0.5\linewidth,angle=0]{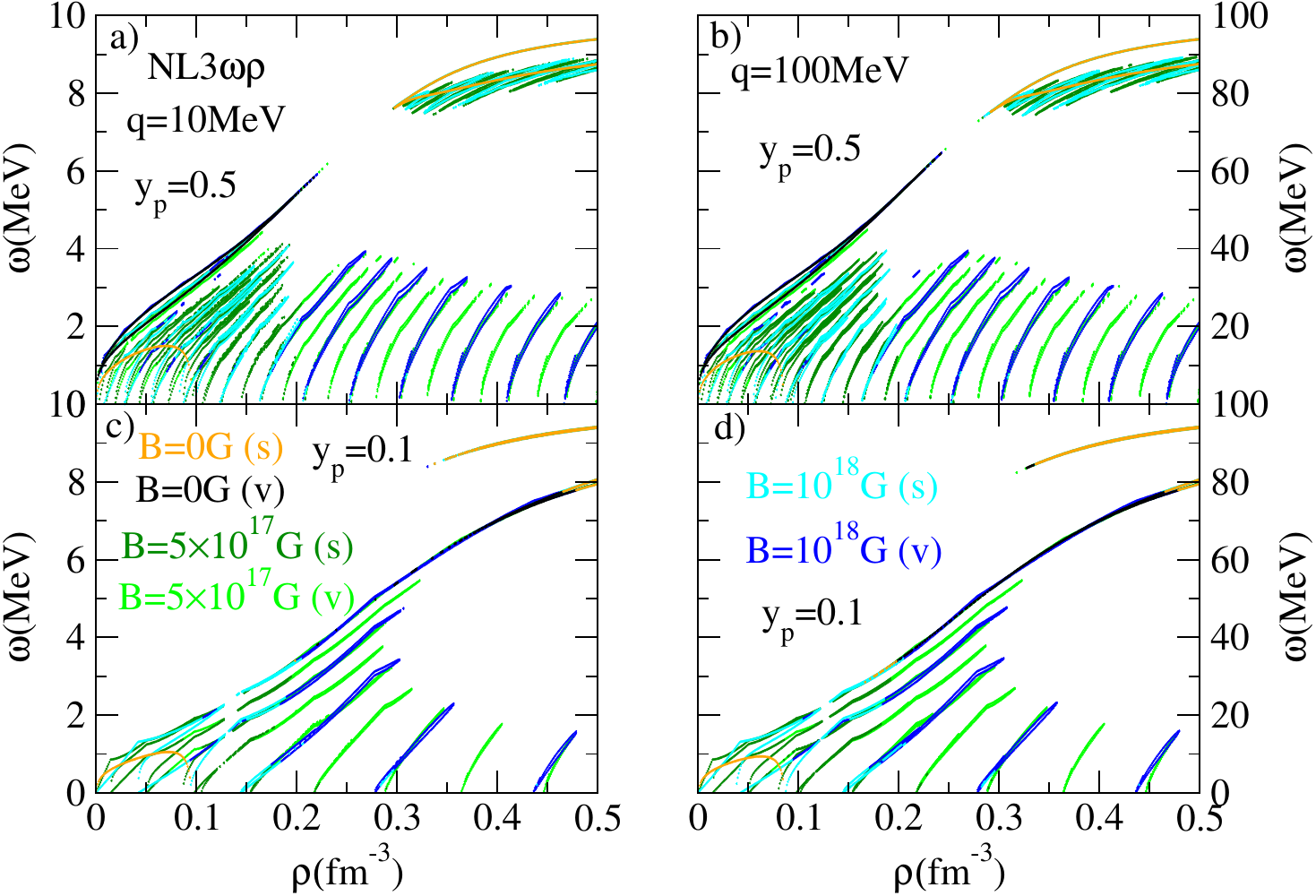}&
\includegraphics[width=0.5\linewidth,angle=0]{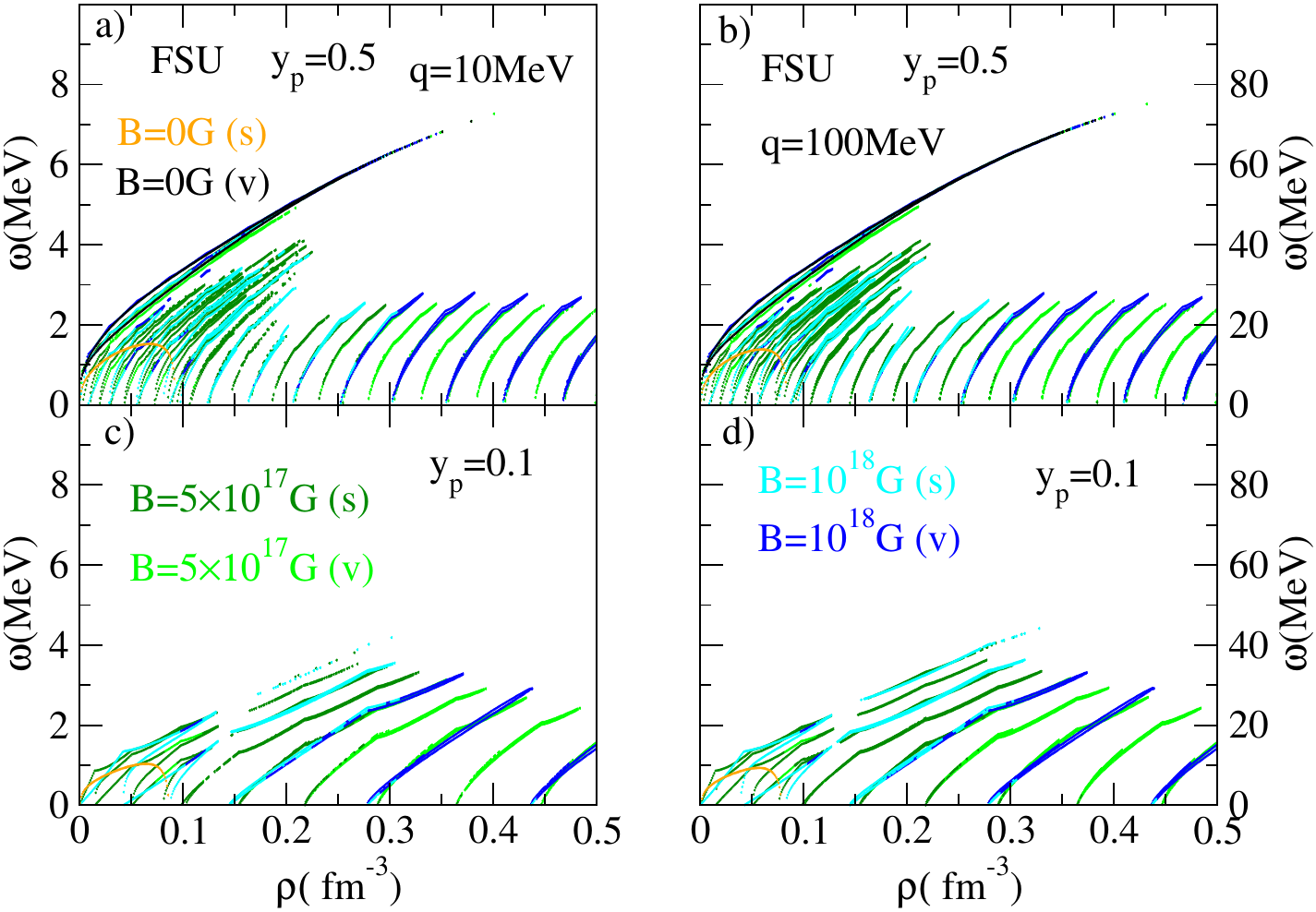}
\end{tabular}
\caption{(Color online) Nuclear collective modes $\omega$(MeV) as function of the baryon density for NL3 model (top quartets), NL3$\omega\rho$ (bottom left quartet) and FSU (bottom right quartet), for $y_p=0.5$ (top panels in each quartet, a) and b)) $y_p=0.1$ (bottom panels in each quartet, c) and d)), for $q=10$MeV (left panels in each quartet, a) and c)), and $q=100$MeV (right panels in each quartet, b) and d)). For $B=0$~G (black color for isovector mode, orange color isoscalar mode), $10^{17}$~G (red color for isovector mode and gray for isoscalar mode), $5\times10^{17}$~G (green color for isovector mode and dark green for isoscalar mode), $10^{18}$~G (blue color for isovector mode and cyan color for isoscalar mode). The results  are for np matter and do not include the Coulomb interaction.}
\label{figure1a}
\end{figure*}
Nuclear collective modes are of particular interest in the inner crust/outer core region of neutron stars because they can affect the transport, cooling, and pulsating properties of neutron stars \cite{Baldo:2008pb,Kobyakov:2013eta,Bedaque:2013fja,Kobyakov:2017dbl,avancini05,Baldo:2019rzt}.
In our study, we will restrict ourselves to the density range extending from 0.05 to 0.5 fm$^{-3}$ corresponding to the outer core. We work in the collisionless regime at a temperature equal to zero and will not consider entrainment. 

The present analysis extends our previous study presented in~\cite{rabhi2025}, where the collective modes of nuclear and neutron star matter were investigated without magnetic field. As indicated above, we consider the  models NL3~\cite{nl3}, NL3$\omega \rho$~\cite{Pais16,Horowitz01} characterized by a stiff EOS at high densities due to the absence or small coupling of the $\omega^4$ term; and FSU~\cite{fsu} which includes additional nonlinear or density-dependent couplings that soften the equation of state and symmetry energy. 
These three models are considered as representative because they show distinct trends in the symmetry energy, its slope, and the behavior of the energy density and the pressure as functions of nuclear density. 

In Fig.~\ref{ener_sym}, the symmetry energy and its slope (left panels) and the energy density and pressure (right panels) are depicted as functions of the baryon density for NL3, NL3$\omega\rho$ and FSU. As shown, models  NL3, NL3$\omega\rho$ are characterized by quite stiff EoS at large densities and models NL3$\omega\rho$ and FSU by a soft symmetry energy. Model  NL3 has a large symmetry energy and corresponding slope at saturation density, and in particular, is characterized by a smaller symmetry energy at sub-saturation densities and a larger one above the saturation density.
\begin{figure*}[t]
\begin{tabular}{ccc}
\includegraphics[width=0.33\linewidth,angle=0]{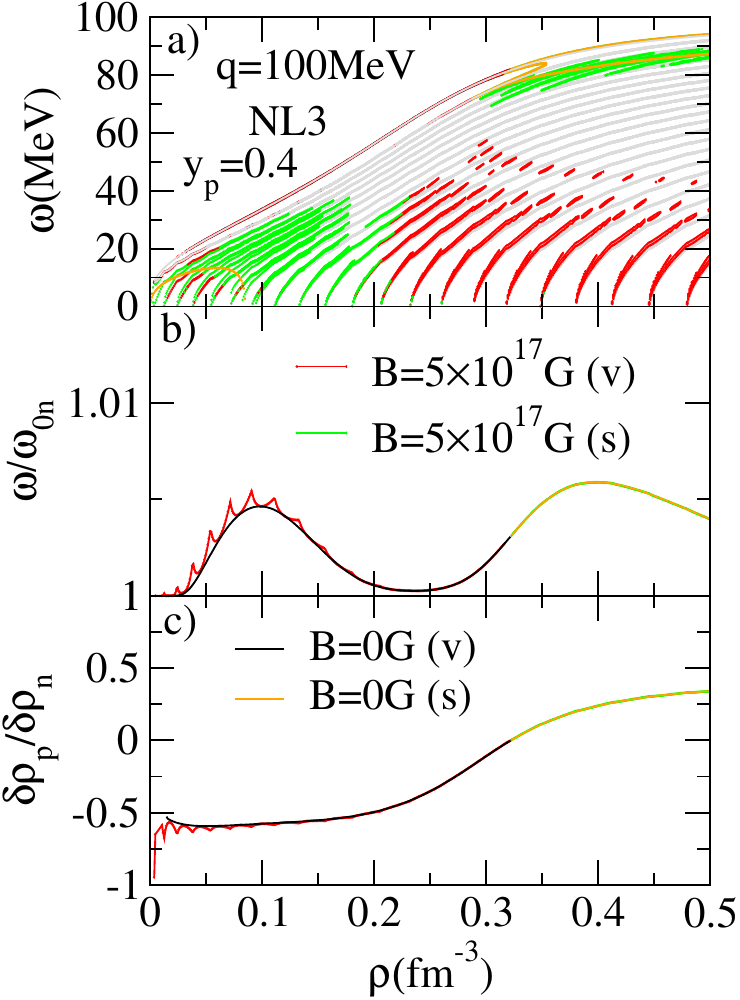} &
\includegraphics[width=0.33\linewidth,angle=0]{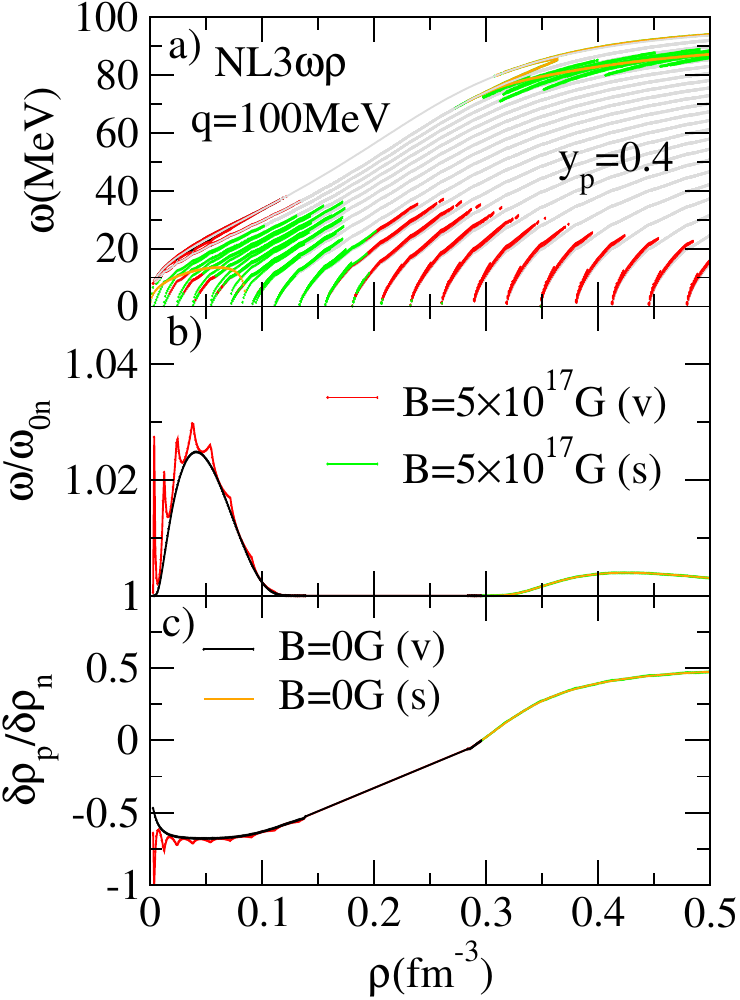} & 
\includegraphics[width=0.33\linewidth,angle=0]{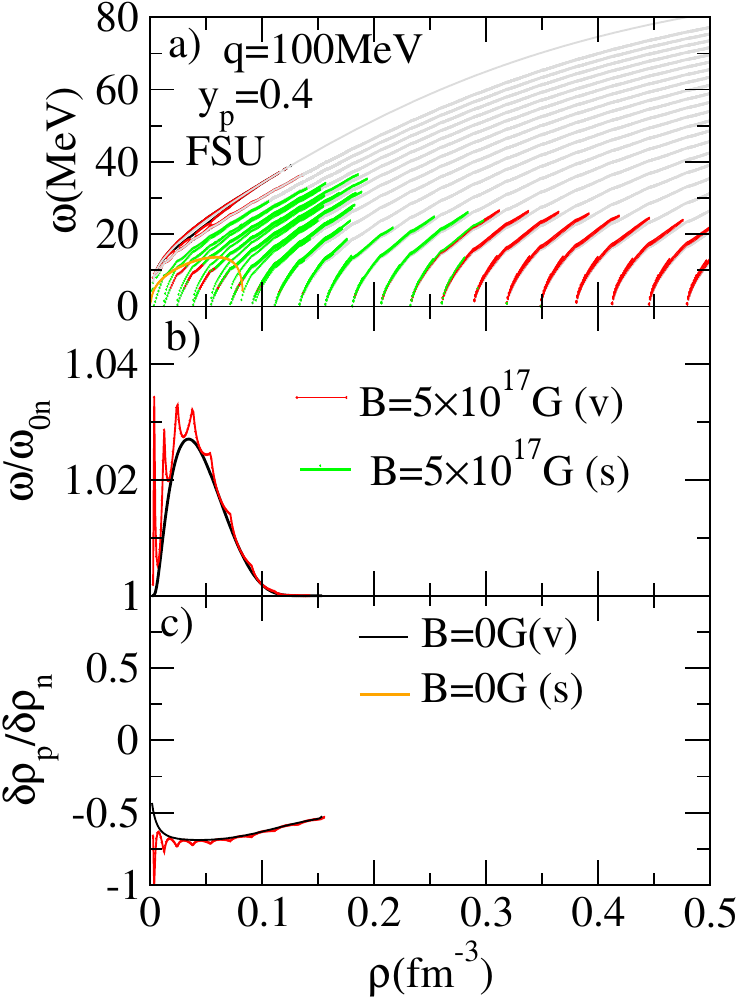}
\end{tabular}
\caption{(Color online) Nuclear collective modes, $\omega$ (MeV), as a function of density for NL3, NL3$\omega\rho$ and FSU models, with proton fraction $y_p = 0.4$ and  for a momentum transfer of $q = 100$~MeV, and magnetic field strengths $B = 0$ and $5 \times 10^{17}$~G. The middle panels of each column show the nuclear collective modes $\omega$ divided by $\omega_{0n}=q V_{F_n}$ versus density, and the bottom panels the amplitude ratio $\delta \rho_p/\delta\rho_n$ as a function of the baryon density.
For $B = 0$~G, isoscalar and isovector modes are shown in black and orange, respectively. For $B = 5 \times 10^{17}$~G, the isoscalar mode is shown in green and the isovector mode in red. Results correspond to neutron-proton (np) matter. For reference we show also the nucleons Fermi modes $\omega_{0i}=qV_{Fi}$ $\omega^{n}_{0p}=q \frac{P^p_{F}(n)}{E^p_{F}(n)}$are also indicated by gray lines.}
\label{fignl3}
\end{figure*}

For the NL3, NL3$\omega\rho$, and FSU models, we present the longitudinal isoscalar and isovector modes $\omega$ as a function of the nuclear density in Fig.~\ref{figure1a} divided into four quartets of four plots. The top left and the two bottom quartets have the results for magnetic field strengths: $B = 0$, $5 \times 10^{17}$ G, and $10^{18}$ G, and on the anti-clock direction correspond, respectively, to NL3, NL3$\omega\rho$ and FSU models. The top right quartet corresponds to $B = 0$ and $10^{17}$ G for the NL3 model, and is representative of what is expected for this smaller magnetic field intensity for the other models.
In each quartet,  panels (a) and (b) show the results for symmetric nuclear matter with a proton fraction $y_p = 0.5$, for momentum transfers of $q = 10$ MeV and $q = 100$ MeV, respectively. These two values have been selected as representative: $q=10$~MeV corresponds to a long wavelength, with a strong contribution of the Coulomb field. If electrons are included in the calculation, they couple more strongly to protons for small values of $q$. 
For $q\sim$ 100 MeV, corresponding to a short  wavelength,  the response of the system is essentially defined by nuclear interactions. If electrons are included, the Coulomb interaction between protons and electrons is weak and the two particles move in a decoupled manner. Panels (c) and (d) correspond to asymmetric nuclear matter with $y_p = 0.1$, a typical proton fraction in the outer core of neutron stars, for the same values of $q$. We adopt the following color scheme to distinguish between the isoscalar and isovector modes at a given magnetic field strength: for $B = 0$ G, the isoscalar mode is shown in black and the isovector mode in orange; for $B = 10^{17}$ G, gray is used for the isoscalar mode and red for the isovector mode; for $B = 5 \times 10^{17}$ G, dark green represents the isoscalar mode and green the isovector mode; and for $B = 10^{18}$ G, the isoscalar mode is shown in cyan and the isovector mode in blue. All results correspond to neutron-proton (np) matter.
\begin{figure*}[t]
\begin{tabular}{cc}
\includegraphics[width=0.5\linewidth,angle=0]{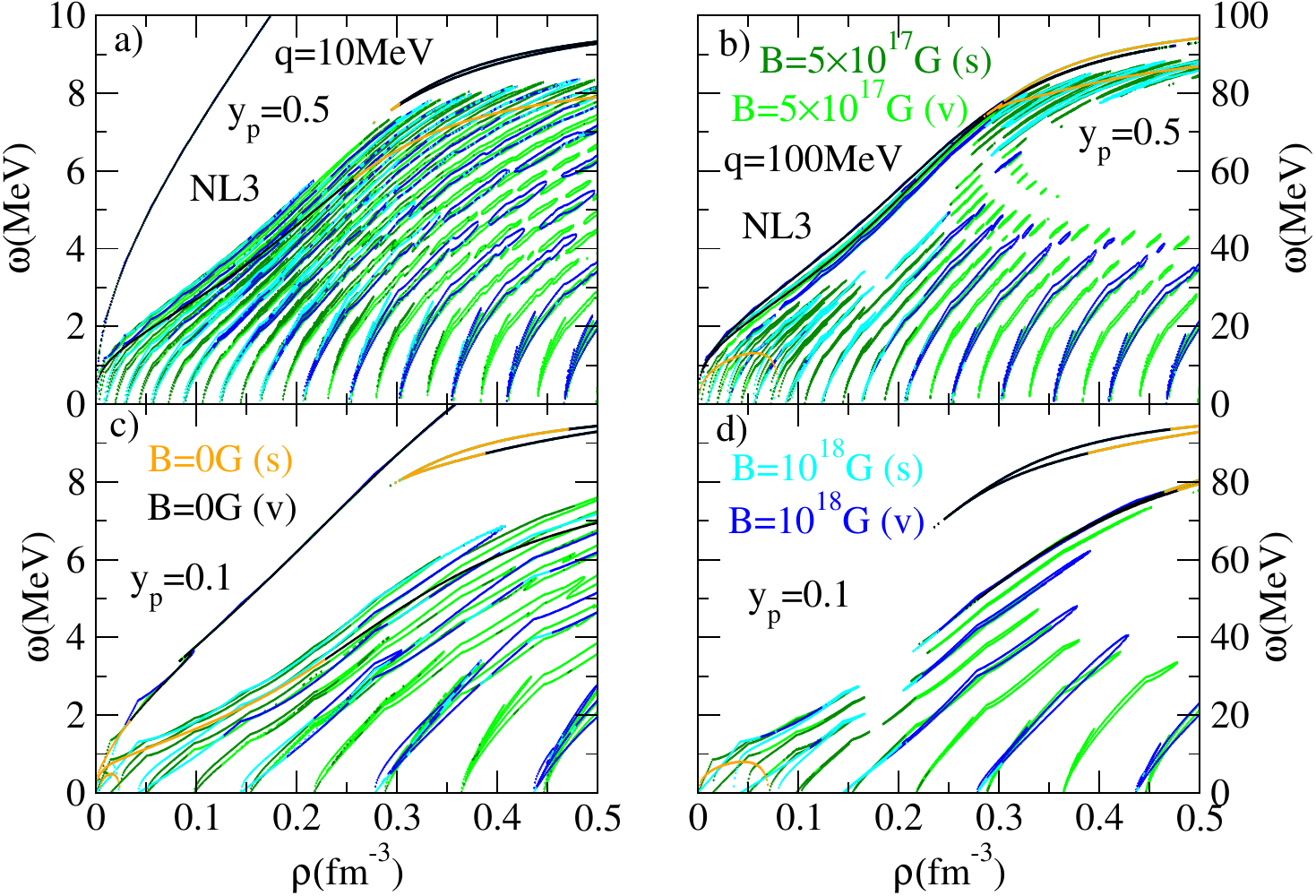}&
\includegraphics[width=0.5\linewidth,angle=0]{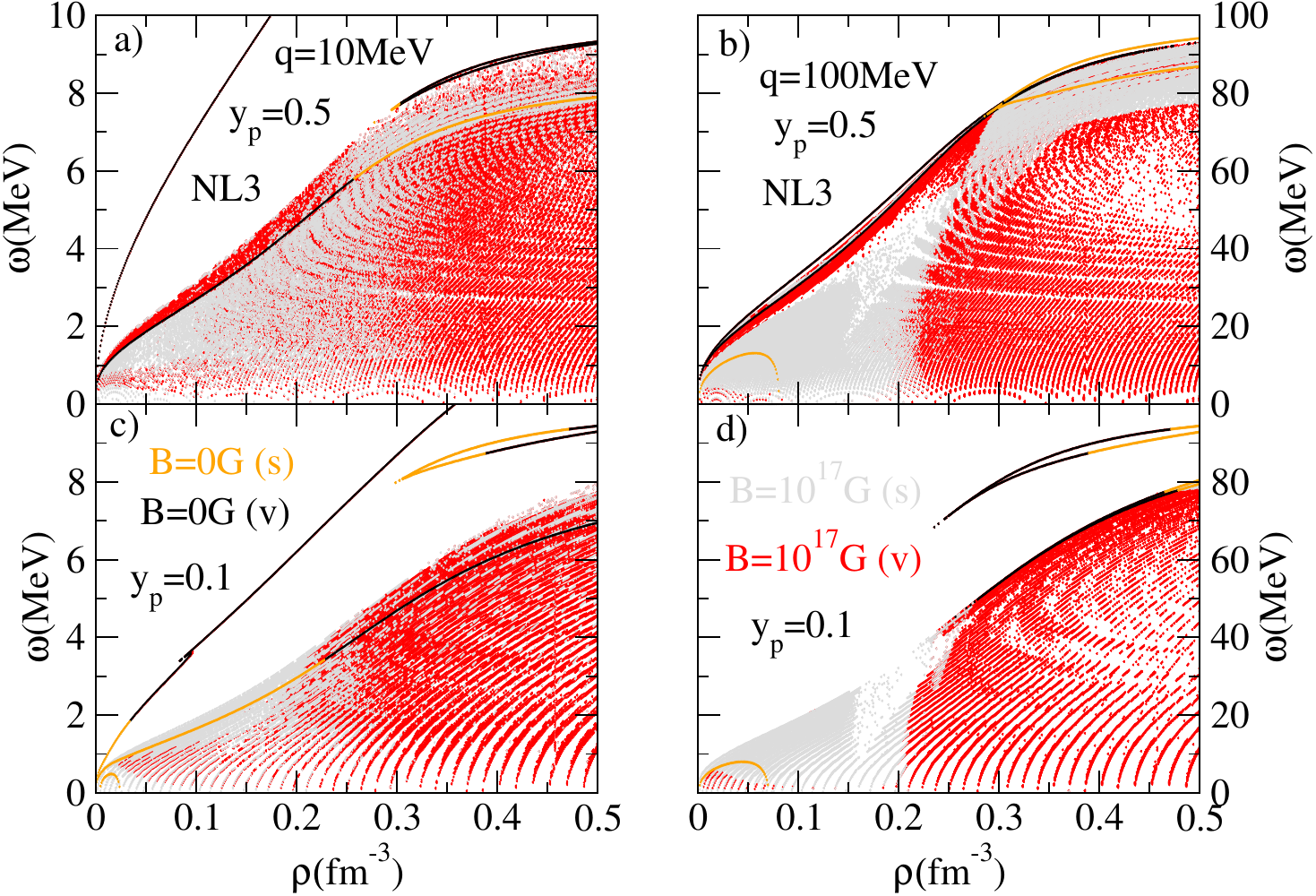}\\
\includegraphics[width=0.5\linewidth,angle=0]{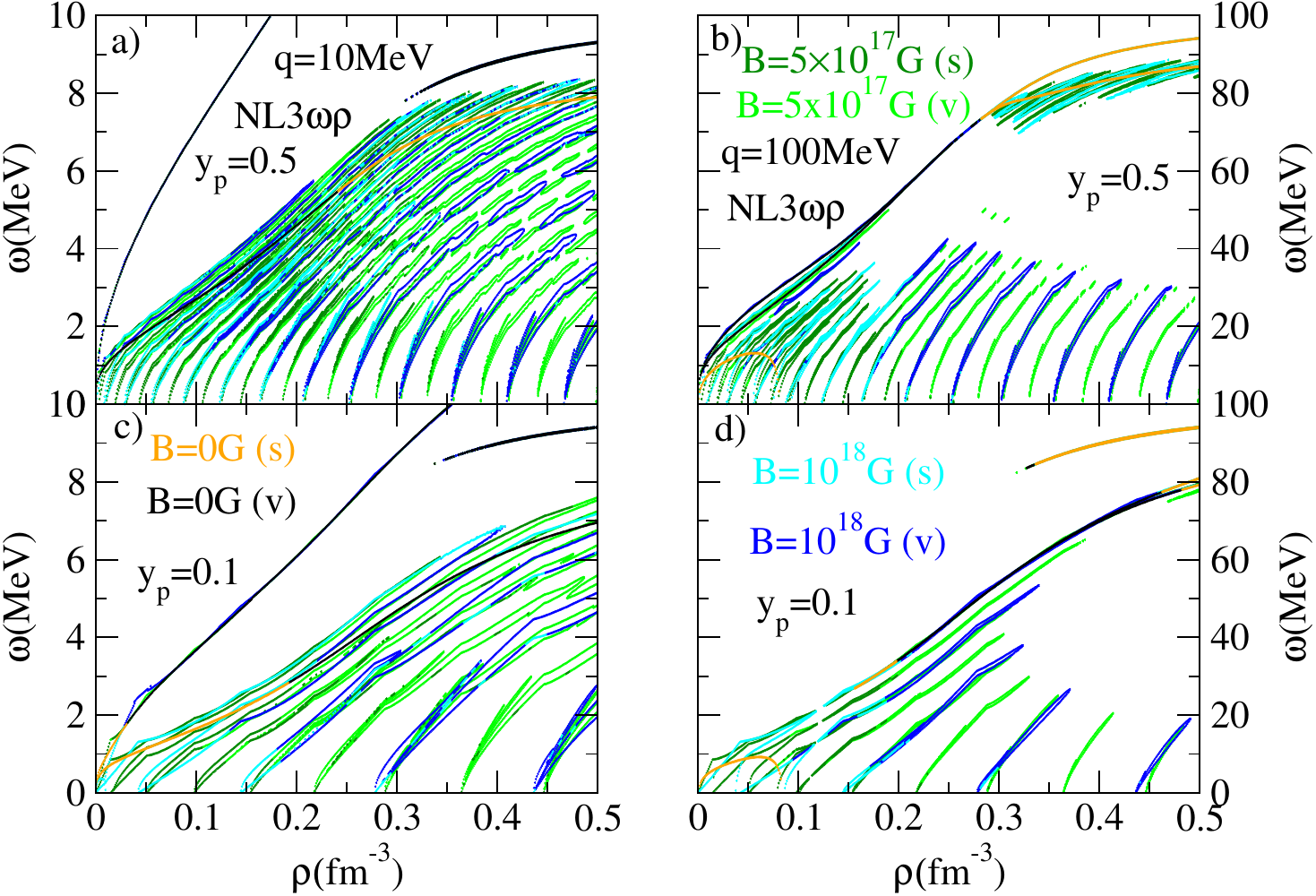}&
\includegraphics[width=0.5\linewidth,angle=0]{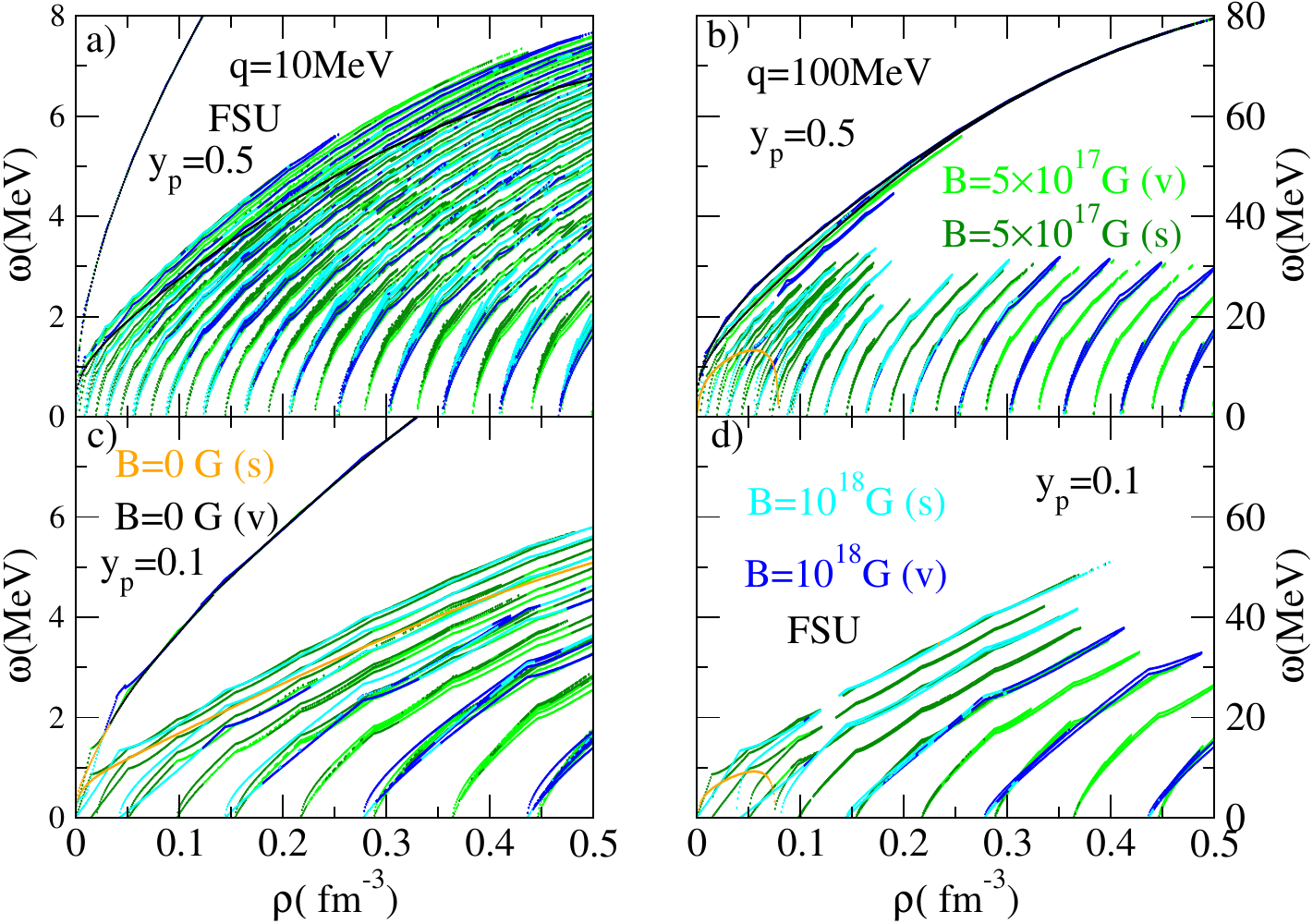} 
\end{tabular}
\caption{(Color online) 
Nuclear collective modes $\omega$(MeV)  as function of the density for NL3 model (top quartets), NL3$\omega\rho$ (bottom left quartet) and FSU (bottom right quartet), for $y_p=0.5$ (top panels in each quartet,  a) and b)) $y_p=0.1$ (bottom panels in each quartet, c) and d)), for $q=10$MeV (left panels in each quartet, a) and c)), and $q=100$MeV (right panels in each quartet, b) and d)). For $B=0$~G (black color for isovector mode, orange color  isoscalar mode), $10^{17}$~G (red color for isovector mode and gray for isoscalar mode), $5\times10^{17}$~G (green color for isovector mode and dark green for isoscalar mode), $10^{18}$~G (blue color for isovector mode and cyan color for isoscalar mode). The results are for np matter and including Coulomb effect.}
\label{figure2a}
\end{figure*}

At zero magnetic field B=0~G, the longitudinal collective excitation spectrum in symmetric nuclear matter distinguishes two classes of RMF models. The first class, which includes NL3 and NL3$\omega\rho$, exhibits three families of modes: (i) a low-density isoscalar mode associated with spinodal instability, (ii) a pair of isovector-like modes at intermediate densities, and (iii) a pair of isoscalar-like modes at higher densities. Each pair consists of a damped mode, lying within the particle-hole continuum, and an undamped mode, located above the Landau damping threshold and thus capable of propagating as a collective excitation. The high-density isoscalar modes are supported by the stiff equation of state, such as NL3 and NL3$\omega\rho$. In contrast, the second class, represented by the FSU model, displays only the low-density isoscalar instability mode and the pair of isovector-like modes, with no high-density isoscalar branch because it is a soft EOS with small incompressibility.

\begin{figure*}[t]
\begin{tabular}{ccc}
\includegraphics[width=0.33\linewidth,angle=0]{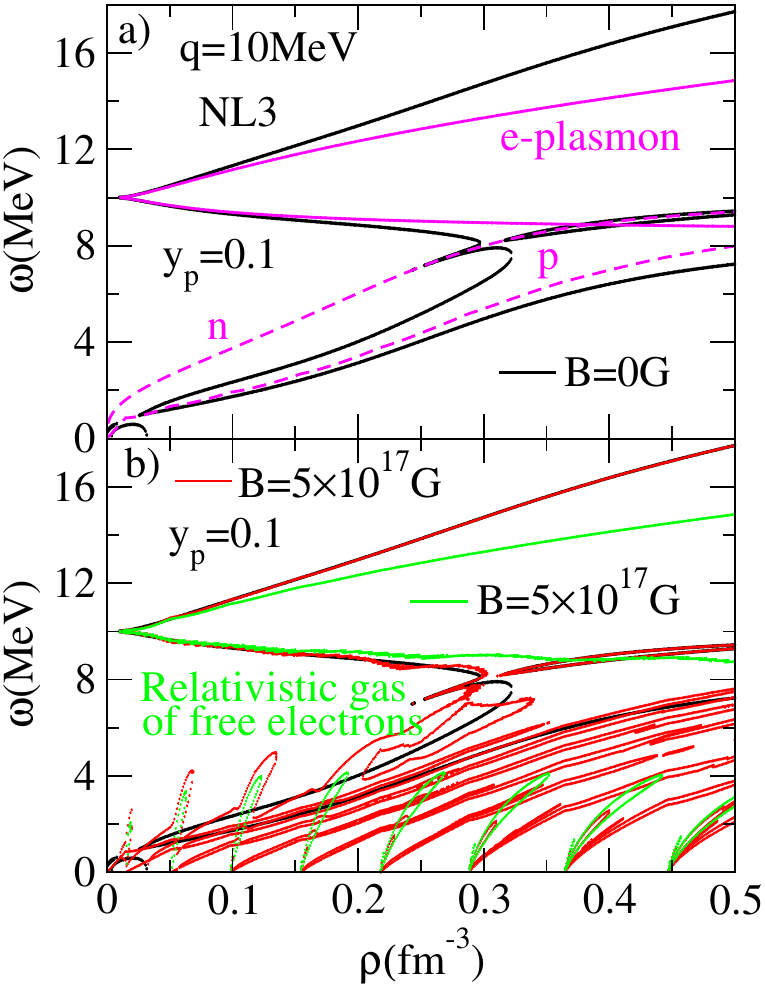}&
\includegraphics[width=0.33\linewidth,angle=0]{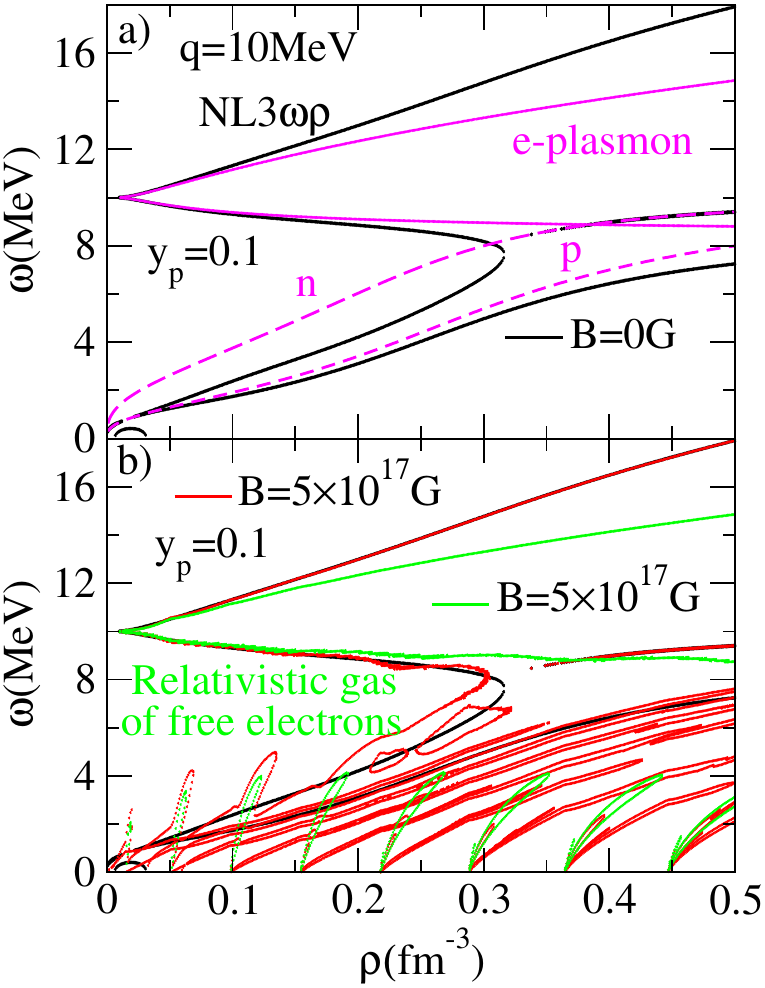}&
\includegraphics[width=0.33\linewidth,angle=0]{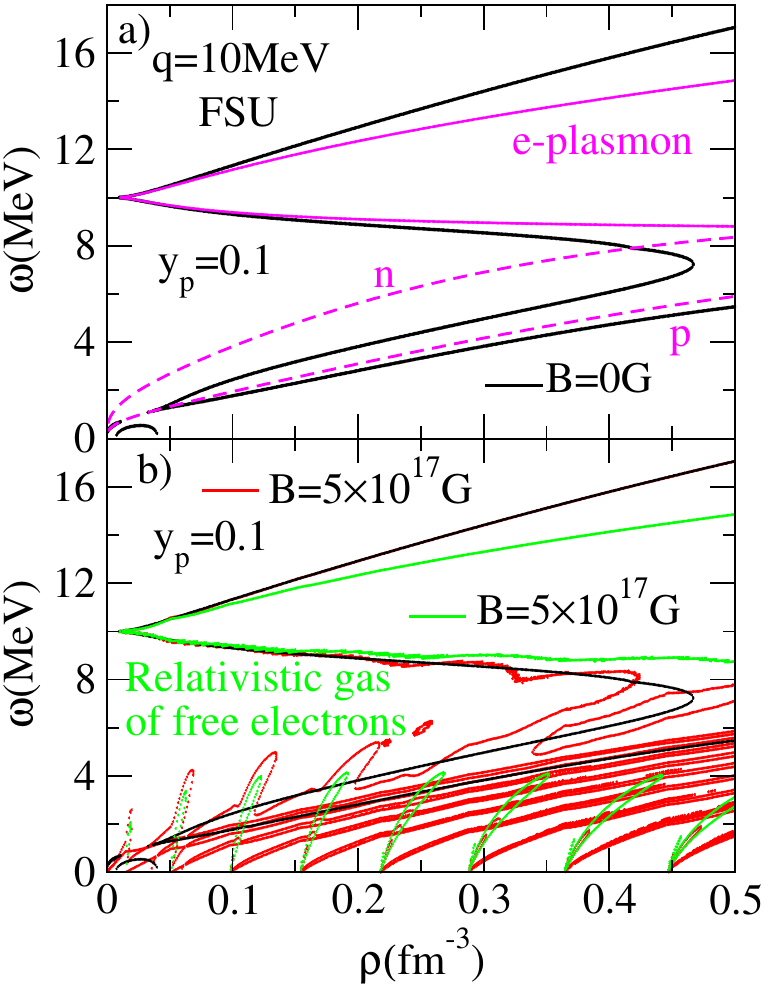}\\
\end{tabular}
\caption{(Color online) Collective modes, $\omega$ (MeV), as a function of the baryon density for NL3 (left), NL3$\omega\rho$ (middle), and FSU (right) models at fixed proton fractions $y_p = 0.1$ and momentum transfer $q=10$~MeV. Two magnetic field strengths are considered: $B = 0$ and $5 \times 10^{17}$~G. The top panels display the results for $B=0$G, where the solid black lines correspond to npe matter, and the solid magenta lines represent the collective modes of a relativistic free electron gas. The nucleons Fermi modes $\omega_{0i}=qV_{Fi}$ are also indicated by magenta dashed lines. The bottom panels compare the collective modes in npe matter for $B = 0$~G (black lines) and  $B = 5 \times 10^{17}$~G (red lines). For reference, the collective modes of a relativistic gas of free electrons for $B = 5 \times 10^{17}$~G are shown in green solid lines.
}
\label{fignl3_npe}
\end{figure*}

When a finite magnetic field is introduced, Landau quantization alters the phase space of charged particles, particularly protons, by discretizing their transverse motion. This modifies the damping conditions of modes associated with proton oscillations and allows the undamped modes to propagate in regimes where they were previously suppressed. Note that neutron-like modes, the high-lying modes, are not affected. In particular, in the FSU model, the presence of a finite magnetic field leads to the emergence of high-density undamped longitudinal modes, despite their absence at B=0~G. This behavior arises due to Landau-level quantization, which enables collective propagation. Each Landau level has an associated Fermi level, and undamped excitations propagate above the Fermi levels. The modes appear by pairs; the mode with the larger energy is the undamped one. The larger the magnetic field, the larger the density range between modes, due to a smaller number of LL. Reducing the strength of the field from $10^{18}$ to $5\times 10^{17}$, the number of modes duplicates and from $10^{18}$ to $10^{17}$ the number of modes increases by almost an order of magnitude (see top right quartet). These findings underscore the significant role of the magnetic field in reshaping the collective mode spectrum and highlight the interplay between EOS properties and magnetic quantization effects in dense nuclear matter. The presence of a strong magnetic field gives rise to a series of low-energy modes that for $B=0$ do not propagate.

The analysis of collective mode frequencies $\omega$ as a function of density for transferred momenta $q$ at $q=10$~MeV and $q=100$~MeV reveals a linear dependence of $\omega$ on $q$, characteristic of zero-sound-like modes and consistent with predictions from the Landau-Fermi liquid theory. 

To further analyze the results presented in Fig.~\ref{figure1a}, we now explore the effect of the magnetic field on stable (undamped) nuclear collective modes with a speed of sound that exceeds the neutron Fermi speed. We consider nuclear matter at a fixed proton fraction $y_p=0.4$.
In Figs.~\ref{fignl3} (top panels), the collective nuclear modes, $\omega$ (in MeV), are plotted as a function of density with proton fractions $y_p = 0.4$, at a momentum transfer of $q = 100$ MeV, and magnetic field strengths $B = 0$ and $5 \times 10^{17}$ G. The middle panels show the sound velocity of the collective modes defined by the ratio $\omega/\omega_{0n}$, where $\omega_{0n}=q V_{F_n}$, as a function of density, allowing for a direct comparison with the neutron Fermi velocity, and the bottom panels show the amplitude ratio $\delta \rho_p/\delta\rho_n$ as a function of the baryon density. For $B = 0$ G, the isoscalar and isovector modes are shown in black and orange, respectively. For $B = 5 \times 10^{17}$ G, the isoscalar mode is shown in green and the isovector mode in red. The isoscalar-like neutron modes, where neutrons and protons oscillate in phase, are only weakly influenced by the magnetic field, since the proton modifications tend to average out and the dominant scalar and vector meson fields are less sensitive to the isospin asymmetry induced by the magnetic field. In contrast, the isovector-like neutron modes, characterized by out-of-phase oscillations with protons, are more sensitive to the magnetic field, as the proton response, which is strongly altered by Landau quantization, feeds back through the isovector channel.
In the absence of a magnetic field and for a proton fraction of $y_p=0.4$, the solutions of the dispersion relation yield two neutron-like and two proton-like modes. These modes can be either Landau damped or undamped, depending on their location relative to the Fermi mode. Additionally, at low densities, an unstable mode emerges, associated with the well-known mechanical (spinodal) instability of asymmetric nuclear matter.
When a magnetic field is introduced, the proton modes are strongly affected by Landau quantization, as already noted, giving rise to the emergence of multiple proton-like collective modes, each corresponding to a specific Landau level. These modes can exhibit either isoscalar-like or isovector-like character, depending on the phase relationship between proton and neutron oscillations. With the exception of the proton-like mode corresponding to the first LL the remaining higher-LL modes are coupled.

The damped neutron-like isoscalar mode is found to be coupled to the undamped proton-like isoscalar mode associated with the first Landau level, which reflects the dispersion equations given in  Eq.~(\ref{det}) complemented with Appendix B.
The remaining proton-like isoscalar modes, corresponding to higher Landau levels, are mutually coupled. This structure highlights the intricate interplay between proton and neutron degrees of freedom in the presence of a strong magnetic field, and the significant impact of Landau quantization on the collective excitation spectrum of asymmetric nuclear matter.

The distinct behavior observed among the models in Fig.~\ref{fignl3} arises primarily from their underlying properties. In particular, FSU does not exhibit an isoscalar mode at large densities, reflecting its softer equation of state. The differences between NL3 and NL3$\omega\rho$ is due to the softer (stiffer) symmetry energy of the second model above (below) saturation density, a smaller symmetry energy making the propagation of the isoscalar mode at high density softer, \textit{i.e.} with a lower energy. Considering $y_p \le 0.3$ the isovector mode at low densities disappears, and the magnetic field has no effect on the high-density isoscalar modes. The bottom panels show the amplitude ratio $\delta \rho_p/\delta\rho_n$ as a function of the baryon density. If positive, the modes are considered isoscalar since neutrons and protons move essentially in phase. The isovector modes, corresponding to an oscillation of protons and neutrons out of phase, are the ones most affected by the magnetic field.
%

We now focus on the study incorporating the Coulomb interaction within the generalized covariant Vlasov approach, applied to both symmetric and asymmetric nuclear matter using the NL3, NL3$\omega\rho$, and FSU models, as shown in Figs.~\ref{figure2a}, which include the same information represented in Fig.~\ref{figure1a} but with the Coulomb field switched on. 
The Coulomb force introduces a long-range repulsion between protons, which modifies the dispersion relation of collective modes, particularly at low densities and small momentum transfers. For $q=10$MeV, its impact is more pronounced, resulting in a noticeable upward shift in the excitation frequencies of isovector-like modes due to the enhanced restoring force associated with charge oscillations. The proton mode is a plasma-like mode corresponding to proton density oscillations in a uniform electron background. At $q=10$~MeV, the isovector model extends to larger energies up to or even above the $B=0$ proton-like mode identified by the orange line. At higher densities and for larger momentum transfers, such as $q=100$MeV, the Coulomb effect becomes less significant, as the dynamics is increasingly dominated by the strong nuclear interaction. This is clearly seen by comparing the panels obtained with $q=100$ MeV in Fig.~\ref{figure1a} with the corresponding ones in Fig.~\ref{figure2a}. The inclusion of the Coulomb term remains essential for a consistent treatment of charge-dependent effects, particularly for low-momentum transfer and asymmetric matter, where proton and neutron responses differ.

Figs.~\ref{fignl3_npe} are obtained when electrons are also allowed to move for neutron-proton-electron (npe) matter with a fixed proton fraction $y_p=0.1$, calculated at a momentum transfer $q=10$~MeV. The results are presented for the NL3, NL3$\omega\rho$ and FSU models, respectively. The top panels are obtained for $B=0$: the solid black lines correspond to the modes of npe matter, the solid magenta lines represent the collective modes of a relativistic free electron gas, and the magenta dashed lines identify the Fermi modes of nucleons $\omega_{0i} = q\, V_{Fi}$. The bottom panels correspond to npe matter for $B = 0$~G (black lines) and  $B = 5 \times 10^{17}$~G (red lines). For reference, the collective modes of a relativistic gas of free electrons for $B = 5 \times 10^{17}$~G are shown in green solid lines. The main effect of the magnetic field is the opening of new low-energy modes, always in pairs. Some of these are closely connected to the electrons as they follow the low-lying modes of a relativistic gas of electron. All others are isovector modes associated with protons already present in Fig.~\ref{figure2a} when the electrons were considered static. No mode close to the neutron Fermi energy propagates if the FSU model is considered, while it is present in the other two models because they give a stiffer EOS.

\section{Conclusions and outlooks}
In this study, we employed the covariant formulation of the Vlasov equation introduced in~\cite{Heinz} and developed in~\cite{avancini2018} to investigate hadronic matter within the framework of a relativistic mean-field model. Using this formalism, we analyzed the normal modes of a relativistic plasma, with a particular focus on magnetized nuclear matter relevant to the interior of magnetars. Specifically, we derived the dispersion relations for longitudinal density fluctuations in nuclear matter modeled  using a relativistic mean-field approach with constant coupling parameters.

To investigate the impact of the magnetic field on the propagation of nuclear collective modes in a magnetized star, we calculate the stable longitudinal modes using three distinct nuclear models, selected for their stiffness and density dependence of the symmetry energy. We calculate the spectrum for magnetized symmetric and asymmetric np matter both neglecting and taking into account the Coulomb interaction. While for a large momentum transfer the Coulomb interaction has a negligible effect on the spectrum, at small momentum transfer the effect on the isovector modes is strong. Due to the Landau quantization, a pair of modes appears for each Landau level. While the neutron-like mode is not affected by the magnetic field, the proton mode suffers noticeable effects. These low-lying modes have been found both when the electrons are considered as a static background and/or as a dynamic gas. The appearance of these modes occurs for models with both a stiff and a soft symmetry energy.  

The present study was carried out using the covariant Vlasov approach combined with the relativistic mean-field model (RMF), employing the parametrizations of NL3, NL3$\omega\rho$, and FSU of the nonlinear Walecka model (NLWM). While only a subset of models was used, we believe that the main conclusions can be extended to other parametrizations within the same sets. It is worth noting that calibrated density-dependent models show a different dependence of the symmetry energy on the density, and it will be important to consider this framework in future studies of collective modes in relativistic asymmetric magnetized nuclear matter within the covariant Vlasov approach.

The main objective of this work was to investigate the effect of a strong magnetic field on the collective modes of asymmetric nuclear matter at zero temperature. The results are particularly relevant to astrophysical scenarios, where neutral, magnetized matter governs the dynamics of dense environments such as core-collapse supernovae and neutron stars. In these contexts, neutrino interactions play a crucial role, as neutrinos carry out most of the energy. The behavior of plasmon modes, including their potential decay into neutrino–antineutrino pairs, is important for modeling neutrino emission from dense matter. All calculations were performed at zero temperature. The inclusion of the magnetic field and finite-temperature effects could further refine the results within covariant Vlasov approach and will be considered in future studies.

\section*{ACKNOWLEDGMENTS}
The present work was partially supported by Fundação para a Ciência e a Tecnologia (FCT), I.P., Portugal, under the  projects UIDB/04564/2020 (doi:10.54499/UIDB/04564/2020), UIDP/04564/2020 (doi:10.54499/UIDP/04564/2020). S.S.A. acknowledges partial support from CNPq-308963/2023-7 and INCT-FNA.

\appendix
\section{Equations of motion for the meson and eletromagnetic fields}
The equations of motion for the mesons and electromagnetic fields follow
from using the Euler-Lagrangian equations (\ref{mesonlag}):
{\footnotesize{
\begin{eqnarray}
 && \partial_t^2 \phi -\nabla^2 \phi + m_s^2 \phi +  \frac{\kappa}{2} \phi^2 +
 \frac{\lambda}{6} \phi^3 = g_s \sum_{j=p,n} \rho_s^{(j)}  \nonumber \\
 &&  \partial_t^2 V_{\mu} -\nabla^2 V_{\mu} + m_v^2 V_{\mu} + \frac{\xi}{6} V_\nu V^\nu V_\mu +
 2 \Lambda_v b_\nu b^\nu V_\mu
  = g_v \sum_{j=p,n} J^{(j)}_\mu  \nonumber \\
 &&  \partial_t^2 b_{\mu} -\nabla^2 b_{\mu} + m_\rho^2 b_{\mu} +
 2 \Lambda_v V_\nu V^\nu b_\mu
  = \frac{g_\rho}{2} \sum_{j=p,n} \tau_j J^{(j)}_\mu  \nonumber \\ 
  &&  \partial_t^2 A_{\mu} -\nabla^2 A_{\mu} 
  = e( J^{(p)}_\mu - J^{(e)}_\mu) =  \sum_{j=p,e} Q_j J^{(j)}_\mu  \label{eomtot}  \ . 
\end{eqnarray}  }}
Now, we consider small deviations from the equilibrium in the fields, as given in Eq.~(\ref{smalldev}), and perform a Fourier transform, obtaining:
{\footnotesize{
\begin{eqnarray}
 && \left[ -\omega^2 + {\bm q}^{~2} + {\tilde m}_s^2 \right] \delta \phi (\bm{q},\omega)
= g_s \sum_{j=p,n} \frac{2 M_j^{\star(0)}}{(2\pi)^3}\int \frac{d^3 p}{E^{(0)}_j}\delta f^{(j)}
\ , \nonumber \\
 && \left[ -\omega^2 + {\bm q}^{~2} + m_v^2 + \frac{\xi}{6}{V^{(0)}_0}^2 +
             2 \Lambda_v {b^{(0)}_0}^2 \right] \delta V_{\mu} +\frac{\xi}{3} {V_0^{(0)}}^2 \delta V_0 ~ 
\delta_{\mu 0} \nonumber \\ 
&& + 4 \Lambda_v V_0^{(0)} b^{(0)}_0 \delta b_0 ~\delta_{\mu 0}
  = g_v  \sum_{j=p,n} \frac{2}{(2\pi)^3}\int \frac{d^3 p}{E^{(0)}_j} p^\mu \delta f^{(j)} 
                                                            \ ,     \nonumber \\
 && \left[ -\omega^2 + {\bm q}^{~2} + m_\rho^2 +
             2 \Lambda_v {V^{(0)}_0}^2 \right] \delta b_{\mu}  \nonumber \\ 
&& + 4 \Lambda_v V_0^{(0)} b^{(0)}_0 \delta V_0 ~\delta_{\mu 0}
  = \frac{g_\rho}{2}  \sum_{j=p,n} \tau_j \frac{2}{(2\pi)^3}
  \int \frac{d^3 p}{E^{(0)}_j} p^\mu \delta f^{(j)} 
                                                            \ ,     \nonumber \\                                                                                                                        
  && \left[ -\omega^2 + {\bm q}^{~2} \right] \delta A_{\mu} 
   =  \sum_{j=p,e} Q_j \frac{2}{(2\pi)^3}
  \int \frac{d^3 p}{E^{(0)}_j} p^\mu \delta f^{(j)}    \ ,  \label{eommes}
\end{eqnarray} }}
where $\tau_{j}=\pm 1$ is the isospin projection for protons and neutrons, respectively, and 
the effective scalar mass is given by:
{\small{
\begin{equation}
 {\tilde m}_s^2 = m_s^2 + \kappa \phi^{(0)} + \frac{\lambda}{2}{\phi^{(0)}}^2 - g_s^2
 \sum_{j=p,n} d \rho^{(0)(j)}_s \ .
\end{equation} }}
\section{Dispersion relation matrix elements for longitudinal modes}

In this appendix, the explicit expressions for the entries of the matrix associated with the longitudinal collective modes Eq.~(\ref{det}) are shown. These matrix elements follow from substituting the field fluctuations given by Eqs.~(\ref{eommes}) into Eqs.~(\ref{displongi},\ref{displongisca}) and then obtaining an expression in terms of the particle and scalar densities. 

To derive the dispersion relations for the longitudinal modes using Eqs.~(\ref{displongi}) and (\ref{displongisca}), it is necessary to evaluate $S_j$, defined in Eq.~(\ref{Lind}). This requires computing the derivative of the equilibrium distribution function, given in Eq.~(\ref{wignereq5}), as follows:
{\small{
\begin{eqnarray}
 &&    D_\parallel^{(j)} = \frac{\partial}{\partial p_\parallel} f^{(0)(j)}  = 
  \sum_{n=0}^\infty  
            \left[ L_{n}(2w^2)-L_{n-1}(2w^2) \right]  \nonumber \\ 
 &&\times (-1)^{n}~ e^{-w^2} (-\delta(p_\parallel -p_F^{(j)}(n)) 
 +\delta(p_\parallel +p_F^{(j)}(n)) )
 \ ,   \label{dpar} \nonumber
\end{eqnarray} }}
with $p_F^{(j)}(n)=\sqrt{\tilde{\mu}_j^2-{M_j^{\star(0)}}^2-2eBn}$ , j=(p,e) and ($g_s=0$) for the electron. From Eq.~(\ref{Lind}), the latter 
equation, and using the property of the Bessel 
function, $J_m(b=0)=\delta_{m 0}$, we find:
{\footnotesize{
\begin{eqnarray}
 && q\; S_j \left[ \frac{J_m^2 (0)}{{E^{(0)}_j}^k}  
  D_\parallel^{(j)} \right] \equiv L_j^{(k)} = \frac{1}{2\pi^2} \sum_{n=0}^{n_{max}} eB
     \int_0^\infty du~  \frac{e^{-u}}{E_{jn}^{(0) k+1}}(-1)^{n+1}   \nonumber \\
     && \times (L_n(2u)-L_{n-1}(2u)) 
 \left( \frac{p_F^{(j)}(n)}{ \left( \frac{\omega}{q}\right)^2 -
 \left(\frac{p_F^{(j)}(n)}{E_{jn}^{(0)}}\right)^2 } \right) \ ,\label{sj}
\end{eqnarray} }}
where $E_{jn}^{(0)}=\sqrt{{M^{\star (0)}_j}^2+{p_F^{(j)}}^2(n)+eBu}$ and notice that we have defined for convenience the generalized Lindhard function $L_j^{(k)}$. The limit $B \to $ of the present formalism is easily obtained since all the information relative to the magnetic field is contained in the generalized Lindhard function, and the latter reduces to the usual Lindhard function in the absence of a magnetic field. The dispersion relations are obtained using the latter equation and Eqs.~(\ref{eommes})
in Eqs.~(\ref{displongi},\ref{displongisca}), yielding:

\begin{eqnarray}
&& \delta \rho^{(i)} + \sum_{j=p,n}\; F^{ij} L_i^{(0)} \delta \rho^{(j)} -  
 \frac{Q_j\;Q_e}{e^2}\; B_A \; L_i^{(0)} \delta \rho^{(e)}   
\nonumber \\
&& + M^{\star\;(0)} \; B_s \; \sum_{j=p,n} \; L_i^{(1)}\;  \delta \bar{\rho}^{\;(j)}_s \; 
= 0  \; ,
\nonumber \\
&& \delta \bar{\rho}^{\;(i)}_s  + \sum_{j=p,n}\; F^{ij} M^{\star \; (0)} L_i^{(1)} \delta \rho^{(j)} 
\nonumber \\
&& -\frac{Q_j\;Q_e}{e^2}\; B_A \;M^{\star \; (0)} \; L_i^{(1)} \delta \rho^{(e)}   
\nonumber \\
&& + M^{\star\;(0)} \; B_s \; \sum_{j=p,n} \; L_i^{(2)}\;  \delta \bar{\rho}^{\;(j)}_s \; 
= 0 \;\; , \; i =(p,n) \; ,  \label{system_eq_rho}
\end{eqnarray}
where 
\begin{equation}
 F^{ij}=-\left( B_v+\tau_i \tau_j B_\rho + \frac{Q_i Q_j}{e^2} B_A \right) \nonumber
\end{equation}
and
{\small{
\begin{eqnarray}
&& B_v = \frac{1}{2\pi^2}\frac{g_v^2}{\omega^2-\omega_v^2} \left( 1 - \frac{\omega^2}{q^2} \right) \nonumber \\
&& B_\rho = \frac{1}{2\pi^2}\frac{(g_\rho/2)^2}{\omega^2-\omega_\rho^2} 
\left( 1 - \frac{\omega^2}{q^2} \right) \nonumber \\
&& B_s = \frac{1}{2\pi^2}\frac{g_s^2}{\omega^2-\omega_s^2} ~,~B_A = -\frac{e^2}{2\pi^2}\frac{1}{q^2}\ ,
\end{eqnarray} }}
with $\omega_s^2=\tilde{m}_s^2+ \bm{q}^{~2}$, $\omega_v^2=m_v^2+ \bm{q}^{~2}$, 
and $\omega_\rho^2=m_\rho^2+ \bm{q}^{~2}$.
The matrix elements of Eq.~(\ref{det}) follows from the Eqs.~(\ref{system_eq_rho}) and are given explicitly as:  
{\small{
\begin{eqnarray*}
&&a_{11} =1+ F^{pp} {\rm L}^{(0)}_p ~,~ a_{12}= F^{pn} {\rm L}^{(0)}_p ~,~
a_{13} = B_A {\rm L}^{(0)}_p  ~,~    \\
&&a_{14}  =  M^{\star (0)} B_s {\rm L}^{(1)}_p ~, ~
a_{15} =  M^{\star (0)} B_s {\rm L}^{(1)}_p \\
&& a_{21}=F^{np} {\rm L}^{(0)}_n ~,~a_{22}=1+ F^{nn} {\rm L}^{(0)}_n ~,~ a_{23}=0 ~,~ \\
&& a_{24} = M^{\star (0)} B_s {\rm L}^{(1)}_n ~,~ a_{25}= M^{\star (0)} B_s {\rm L}^{(1)}_n \\
&& a_{31} = B_A {\rm L}^{(0)}_e ~,~ a_{32}= 0 ~,~ a_{33}= 1- B_A {\rm L}^{(0)}_e ~,~ a_{34}=0 ~, \\
&&a_{35}=0 \\
&& a_{41} = F^{pp} M^{\star (0)}  {\rm L}^{(1)}_p ~,~ a_{42}= F^{pn} M^{\star (0)}  {\rm L}^{(1)}_p ~,~ 
 \\
&& a_{43}= B_A M^{\star (0)} {\rm L}^{(1)}_p ~,~ 
a_{44}=1+ B_s {M^{\star (0)}}^2 {\rm L}^{(2)}_p ~,~ \\
&& a_{45}=B_s {M^{\star (0)}}^2 {\rm L}^{(2)}_p \\
&&  a_{51} = F^{np} M^{\star (0)}  {\rm L}^{(1)}_n ~,~ a_{52}= F^{nn} M^{\star (0)}  {\rm L}^{(1)}_n 
~,~a_{53}= 0 ~,~ \\
&& a_{54}=B_s {M^{\star (0)}}^2 {\rm L}^{(2)}_n ~,~ a_{55}= 1+ B_s {M^{\star (0)}}^2 {\rm L}^{(2)}_n \ .
\end{eqnarray*} }}
%
%

\bibliographystyle{apsrev4-1}
\bibliography{bibliography}

\end{document}